# Data-Efficient Electromagnetic Surrogate Solver Through Dissipative Relaxation Transfer Learning

Sunghyun Nam[1], Chan Y. Park[2*], and Min Seok Jang[1*]

[1]School of Electrical Engineering, Korea Advanced Institute of Science and Technology, Daejeon 34141, Republic of Korea

[2]KC Machine Learning Lab, Seoul 06181, Republic of Korea

*Corresponding author.

Email: Sunghyun Nam: felix0616@kaist.ac.kr, Chan Y. Park: chan.y.park@kc-ml2.com,
Min Seok Jang: jang.minseok@kaist.ac.kr

## Abstract

In neural network surrogate solvers for electromagnetic simulations, accurately modeling resonant phenomena remains a central challenge. High-amplitude resonances generate strongly localized field patterns that deviate significantly from the general distribution of non-resonant cases, leading to instability and degraded predictive performance. To address this, we introduce dissipative relaxation transfer learning (DIRTL), a data-efficient training framework that integrates transfer learning with loss-regularized optimization principles from high-Q photonics. DIRTL first pretrains the model on data generated with a small fictitious material loss, which broadens sharp resonant modes and suppresses extreme field amplitudes. This smoothing of the response landscape enables the model to learn global modal features more effectively. The pretrained model is subsequently fine-tuned on the target lossless dataset containing true high-amplitude resonances, allowing stable adaptation based on the pretrained representation. Applied to both the Fourier Neural Operator (FNO) and UNet architectures, DIRTL yields substantial improvements in prediction accuracy, including up to a two-fold error reduction for the FNO variant. Furthermore, DIRTL demonstrates robustness across diverse training conditions and supports multi-tasking performance, suggesting the generalizability and flexibility of the pretrained core. Altogether, these results position DIRTL as a physically grounded curriculum for improving the reliability of neural network surrogate solvers.

## Table of Contents Graphic

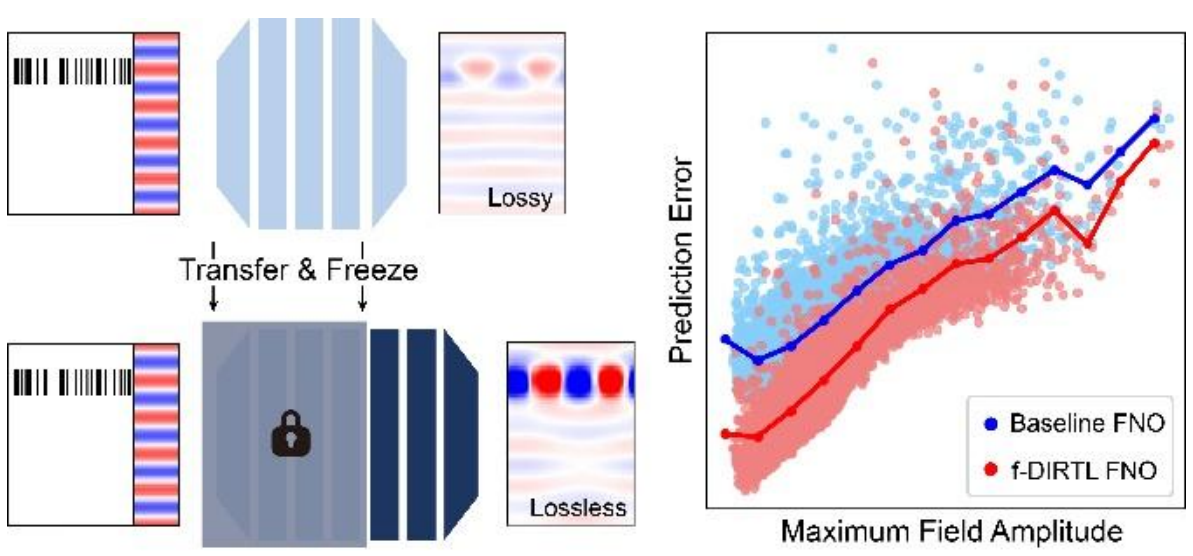

**For Table of Contents Only**

## 1. Introduction

In recent years, machine learning (ML)-based approaches have been widely adopted in nanophotonics research, ranging from alternative device-optimization strategies using generative models and reinforcement learning to accelerating conventional inverse-design workflows. In particular, ML-driven surrogate solvers have gained significant attention due to their strong potential for integration into modern inverse-design pipelines. Conventional numerical methods for electromagnetic (EM) simulations, such as the finite-difference time-domain (FDTD) method and rigorous coupled-wave analysis (RCWA), provide accurate solutions but are often computationally expensive, thereby becoming a major bottleneck in iterative design processes. To address these limitations, neural network (NN)-based surrogate models have been explored to accelerate simulations and enable high-speed exploration of complex design spaces [1-4].

In training NN-based surrogate solvers for EM simulations, a significant challenge arises when dealing with resonant structures. Owing to the spectral bias of NNs, models preferentially learn smooth, low-frequency features during training, making it inherently difficult to capture the sharp spatial variations and large gradients associated with resonant responses [5,6]. As a result, resonant field patterns, characterized by strong localization and high spatial frequency components, are underrepresented. To achieve accurate forward representations, this limitation must be addressed within surrogate modeling frameworks, where precise inference of the underlying Maxwell equation solutions is essential. Such accuracy is critical for reliable downstream inverse-design applications, including gradient-based optimization, latent-space search methods, and ML-assisted optimization frameworks [2–4,7–11].

Beyond this intrinsic limitation, the difficulty is further exacerbated by the dataset's statistical nature. Highly resonant cases are typically scarce and deviate significantly from the dominant background scattering patterns, effectively acting as outliers [12-15]. Consequently, even when the model attempts to fit these cases, their high variance and sensitivity make them difficult to learn consistently, leading to underfitting of resonant responses. At the same time, the presence of such outliers can distort the overall learning process, resulting in overfitting behaviors where non-resonant cases are incorrectly predicted as resonant. Most prior work on NN-based PDE solvers has focused on mitigating the impact of such outlier cases [16,17]. However, this strategy is not suitable for EM problems, where resonant modes play a central role in the functionality and design of photonic devices [18,19].

The majority of existing approaches for improving NN surrogate solvers primarily focus on model design, modifying network architectures to enhance generalization and expressivity, thereby enabling them to handle more diverse and complex physical systems. For accurate field predictions, NN models such as DeepONet [20, 21], UNet [7, 14, 22-27], and the Fourier Neural Operator (FNO)[12,15,28-30] have been extensively explored. While these strategies have demonstrated notable performance gains, they remain largely problem-specific, with solutions often tailored to specific datasets or device configurations. More critically, although such methods can alleviate certain challenges in photonics modeling, they do not directly address the intrinsic difficulty of accurately capturing resonant behaviors that dominate many photonic systems.

Another important line of research incorporates physics-based loss functions during learning. In EM systems, this often takes the form of a Maxwell loss, which quantifies how well the surrogate solver adheres to Maxwell's equations. By embedding this physical knowledge directly into the training objective, the model is encouraged to learn physical behavior rather than merely interpolating data, thereby improving data-efficiency and generalization [15, 23-26]. However, a key challenge remains in that highly resonant responses are difficult to capture, even for physics-informed approaches, due to their multi-scale nature. In resonant cases, the field consists of a smoothly varying background overlaid with sharply localized, high-amplitude features. This combination poses significant challenges during training, as physics-informed neural networks (PINNs) inherit spectral bias, which limits their ability to resolve localized high-frequency features [31, 32]. In addition, PINNs do not directly address the outlier nature of resonant cases. For this, solely addressing the physical aspect will prove insufficient for the inference of resonant features.

In this work, we propose a data-driven training framework that combines conventional photonic optimization principles with transfer learning to address challenges from high-resonant training regimes. Inspired by optimization strategies for high-Q photonic structures, where artificial material loss is introduced to broaden sharp resonances and subsequently removed to recover optimal performance [33-36], we formulate an analogous two-stage learning process termed *dissipative relaxation transfer learning* (DIRTL). This approach enables smooth convergence and enhanced generalization by guiding the network through a refined learning landscape.

Consequently, DIRTL achieves substantially improved prediction accuracy with approximately nine-fold sample efficiency for the given problem set, demonstrating its effectiveness as a data-efficient framework for EM surrogate solvers. We further propose that DIRTL's ability to enable robust and data-efficient surrogate model training can benefit inverse-design workflows by allowing more accurate evaluation of resonant responses. This improved fidelity, combined with reduced data requirements, enhances the practicality of surrogate solvers in large-scale and iterative design settings.

## 2. Dissipative Relaxation Transfer Learning

### 2.1 Problem Description

As a physical problem, we consider a one-dimensional (1D) multi-wavelength binary grating, which is described in **Fig. 1(a)**. The grating consists of a $1 \times 64$ array in free space, illuminated by a transverse magnetic (TM) plane wave normally incident on the device. Each element alternates between free space and a dielectric material with refractive index $n = 2$. The training and testing datasets span wavelengths from 650 nm to 750 nm in 10 nm increments, with all EM field data generated using the `meent` RCWA solver [37]. Simulations employed 81 Fourier orders, which were shown to be sufficient for numerical convergence (see **Section S1 of the Supporting Information**).

From this problem setup, we observe the emergence of the resonant outlier cases as described above. As illustrated in **Fig. 1(b)**, most of the generated test samples are non-resonant, not showing pronounced localized field enhancements in the grating space, while cases of highly localized fields with large field amplitudes occasionally emerge. This distribution is quantified in **Fig. 1(c)**, displaying a long tail corresponding to strongly resonant cases with maximum amplitudes reaching over 50 times the incident field amplitude.

As the baseline surrogate solver, we employed the FNO model [28] consisting of a lifting layer, a series of Fourier layers, and a projection layer, which has been widely adopted in recent electromagnetic simulation studies. Both lifting and projection layers consist of $1 \times 1$ pointwise convolutions, mapping the input in and out of the iterative Fourier layers. To further examine the general applicability of our training strategy, we also implemented a UNet model, which has demonstrated comparable effectiveness in electromagnetic field prediction and inverse-design tasks [22]. UNet follows the standard encoder–decoder architecture with residual connections, which is commonly used in image segmentation (see **Section S2 of the Supporting Information**).

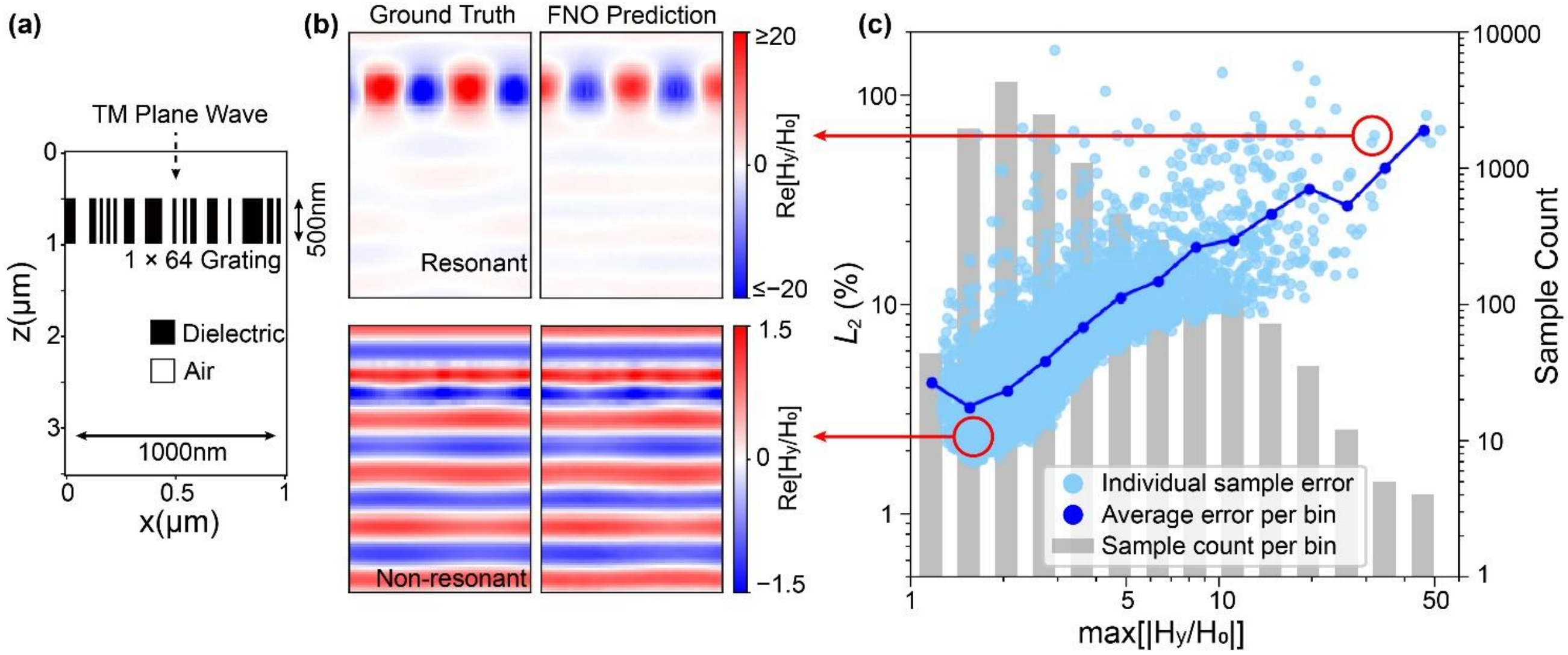

**Figure 1 Problem schematic for DIRTL testing**. (a) Data generation setup. Training and test samples are produced by randomly generating binary grating structures. (b) Representative field distributions for resonant and non-resonant cases. Resonant structures exhibit strongly localized, high-amplitude fields concentrated within the grating region, whereas non-resonant cases show broadly distributed, weakly varying fields. (c) Baseline FNO performance trained only on lossless target data with the bin-averaged error (blue markers) and the corresponding data-distribution histogram (gray bars). Each light-blue dot indicates the prediction error versus the maximum field amplitude for a single test sample, and blue markers denote the average error within logarithmically spaced amplitude bins. The histogram shows an exponential decrease in sample count with increasing maximum amplitude, with extreme resonant cases (amplitudes approaching 50) comprising less than 0.1% of the dataset. Red circles highlight the locations of the resonant and non-resonant examples from panel (b) within the overall error–amplitude distribution.

To incorporate wavelength information effectively, we adopted the wave prior approach introduced in NeurOLight [29]. In this method, the wavelength is encoded as a sinusoidal function that mirrors the incident plane wave, providing both explicit wavelength information and a strong initial guess for the model to refer from. For each training sample, both the device structure and the corresponding wave prior were used as model inputs. The networks were trained using the mean squared error (MSE) loss, while optimized by the Adam optimizer [38]. The learning rate was tuned through Bayesian optimization with a step learning rate scheduler, which provided the most stable and efficient convergence. The detailed hyperparameters for both models and learning rates are summarized in **Section S2 of the Supporting Information**.

When trained on such data, the baseline FNO model displays an order of magnitude higher prediction errors in regions associated with strong resonances as shown in **Fig. 1(c)**. Furthermore, the model's effort to fit these extreme resonant responses can lead to overfitting, which reduces prediction accuracy for non-resonant structures (see **Section S2 of the Supporting Information**). These behaviors highlight the intrinsic difficulties in EM surrogate solvers due to highly resonant responses.

## 2.2 High-Q Optimization in Traditional Photonics

In conventional photonic design, optimizing high-Q resonant features is intrinsically difficult. The extremely narrow spectral linewidths associated with high-Q modes create sharply peaked, highly nonconvex objective landscapes, making it challenging for optimization algorithms to accurately locate optimal configurations [39,40]. A widely used remedy is to introduce fictitious material absorption during optimization. This artificial loss broadens narrow spectral features, which smooths the overall landscape. From this, it becomes easier to capture the relevant modal behavior, enabling more reliable exploration of the global design space. Once a promising configuration is found, the added loss can be gradually removed, allowing convergence towards the true high-Q response [33-36].

This effect can be understood through the quasinormal mode (QNM) framework. QNMs, which are solutions to the sourceless Maxwell equations, define the resonant modes of an EM system. QNMs of a system consist of eigenmodes $\tilde{\mathbf{E}}_m$ and corresponding eigenfrequencies $\tilde{\omega}_m$. Such eigenfrequencies are often mentioned as poles, as they are found as the divergent poles of the system for a certain physical response in the complex frequency plane [18]. Now, given a structure with permittivity distribution $\epsilon(\mathbf{r})$, adding a small imaginary component $i\delta\epsilon(\mathbf{r})$ shifts the QNM poles deeper into the complex frequency plane while leaving the real frequency component essentially unchanged. First-order perturbation theory yields a purely imaginary eigenfrequency correction as below [36].

$$\delta\tilde{\omega}_m = -i\frac{\tilde{\omega}_m \int \tilde{\mathbf{E}}_m^* \cdot \delta\epsilon(\mathbf{r})\tilde{\mathbf{E}}_m dv}{2\int \tilde{\mathbf{E}}_m^* \cdot \epsilon(\mathbf{r})\tilde{\mathbf{E}}_m dv}$$

Such small losses leave the modal field profiles largely intact while not shifting the resonant frequency, but increase the imaginary part of the frequency, which contributes to effective damping of the resonant mode amplitudes (see **Section S3 of the Supporting Information** for detailed derivation).

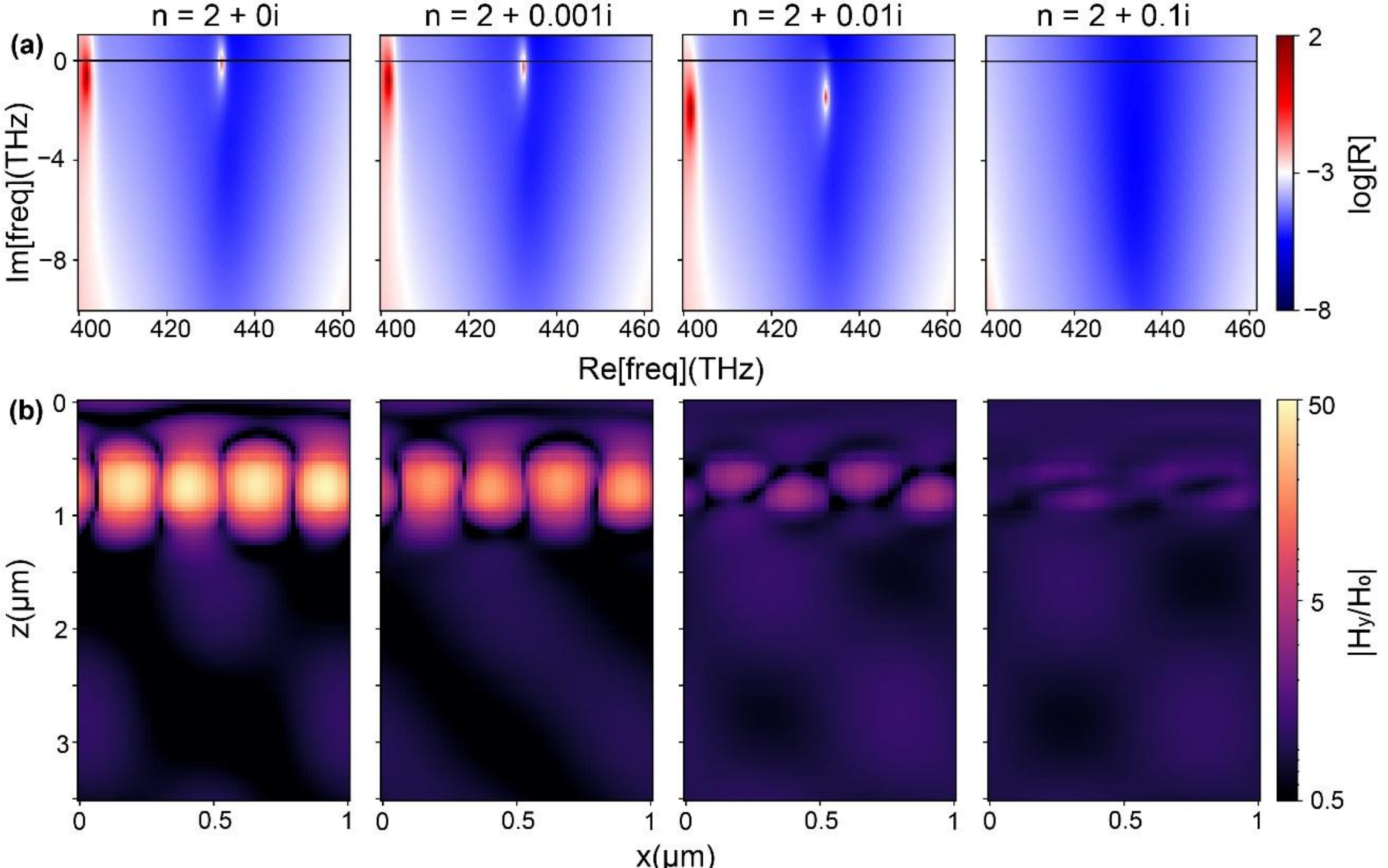

**Figure 2 Relaxation of resonant structures via fictitious material loss.** (a) Complex frequency sweep of device reflectivity. Introducing a small imaginary component to the refractive index shifts the resonance pole near 429 THz (≈700 nm) deeper into the complex frequency plane. The shift occurs almost entirely along the imaginary axis, indicating damping of field amplitudes while leaving the real resonance frequency essentially unchanged. (b) Corresponding field distributions at 700 nm. Adding controlled material absorption attenuates the resonant field intensity within the grating region while preserving the underlying modal profile. Among the tested values, $n = 2 + 0.01i$ provides the most effective balance between amplitude reduction and mode fidelity (see **Section S4 of the Supporting Information**).

### 2.3 Physics-Inspired Transfer Learning Strategy

The same effect can be applied in our electromagnetic surrogate-model training domain. Certain device configurations exhibit strong guided-mode resonances, generating outlier samples in both training and test datasets. Introducing fictitious absorption, which is implemented as an additional dielectric loss term, shifts the corresponding QNMs along the imaginary axis, as illustrated in **Fig. 2(a)**. Following the QNM shift, we see a reduction in amplitudes, which is illustrated in **Fig. 2(b)**.

Borrowing the idea of relaxing resonant field intensities to promote more global optimization behavior, we reinterpret this process within a curriculum learning framework for NN training [41]. Specifically, we design a two-stage learning curriculum that guides the model to first learn general modal features before confronting the full complexity of highly resonant cases. In the first stage, the model is trained on data generated with artificial material loss included (lossy data), where sharp resonances are suppressed. This effectively simplifies the learning landscape, allowing the model to capture smooth, global response characteristics. In the second stage, the pretrained model is transferred and fine-tuned on the target dataset generated without material loss (lossless), which includes high-amplitude resonant configurations. We name this process dissipative relaxation transfer learning, or DIRTL.

For our specific 1D binary grating example, we introduced a small dissipative term to the refractive index ($n = 2 + 0.01i$) chosen to achieve effective amplitude attenuation while preserving the overall modal profiles. The value $0.01i$ added was shown to be most effective among several test values (see **Section S4 of the Supporting Information**). Such values can be heuristically derived via quasinormal mode analysis, which supports our choice of material loss (see **Section S3 of the Supporting Information**). In the baseline configuration, where the model was trained solely on lossless data, 9,000 random structures were used, with field data sampled at each wavelength point (9,000 samples × 11 wavelength points, in total 99,000 training and test samples). In the DIRTL setup, we used 4,500 random structures for both the pretraining (lossy) and fine-tuning (lossless) stages, ensuring that the total amount of training data was equivalent to that of the baseline (49,500 training samples for each stage). The validation and test set consisted of 1,000 random structures generated under lossless conditions, using the same wavelength sampling (11,000 validation and test samples).

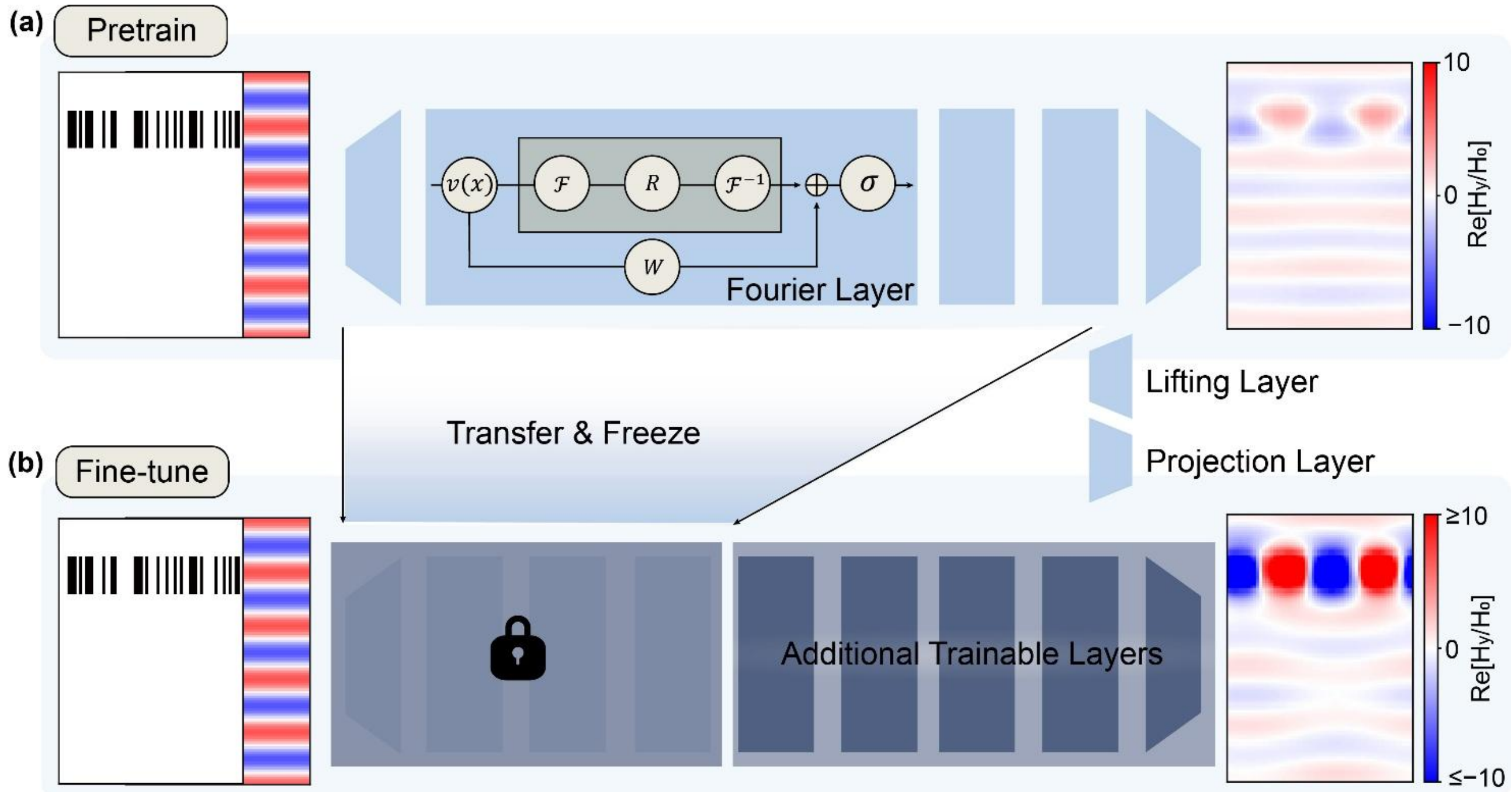

**Figure 3 f-DIRTL FNO training strategy.** (a) Pretraining stage. A 6-layer Fourier Neural Operator (FNO) is trained using structures paired with a wave-prior encoding, which provides both wavelength information and a physically informed ansatz. The pretraining targets are field distributions computed with a small fictitious material loss, resulting in smoothed, attenuated resonances. (b) Fine-tuning stage. The final 10-layer FNO is initialized by transferring and freezing the lifting layer and the first six Fourier layers from the pretrained model. The remaining Fourier layers and the projection layer are then trained on the target lossless dataset, which includes high-Q resonant responses.

We first perform fine-tuning directly from the pretrained model with all parameters unfrozen, which we refer to as l-DIRTL, with l standing for loose. In this case, both the pretraining and fine-tuning stages are conducted using the same 10-layer FNO architecture. However, such an approach can lead to forgetting of the pretrained representations. To mitigate this issue, we introduce f-DIRTL, with f standing for freeze, where a smaller 6-layer FNO is first pretrained on the lossy dataset, and its parameters are subsequently transferred to a larger 10-layer FNO. The transferred layers are frozen to preserve the learned global features, while the additional deeper layers are trained to capture high-amplitude resonant characteristics. For a fair comparison, the baseline, l-DIRTL, and f-DIRTL models all share the same final network architecture. The f-DIRTL FNO schematic is shown in **Fig 3**.

Meanwhile, for a fair comparison with conventional curriculum learning strategies, we evaluated two alternative data-sampling approaches. The first is a data-driven curriculum baseline, in which resonant samples are gradually introduced during training [41]. Specifically, over a total of 100 training epochs, high-amplitude resonant data are incrementally included in the training dataset every 25 epochs, progressing from low to high-amplitude cases. The second approach is an oversampling strategy, where the minority high-Q resonant samples are duplicated within the training dataset to compensate for their inherent scarcity [42]. Detailed implementation procedures and results for both approaches are provided in **Section S5 of the Supporting Information**.

In the case of UNet, both the freeze and loose variants use identical architectures during pretraining and fine-tuning. For f-DIRTL, only the down-sampling path (the encoder that extracts the underlying physical features) is transferred and frozen during fine-tuning, whereas in l-DIRTL the entire network is updated end-to-end (see **Section S4 of the Supporting Information**). For additional baseline comparison, we implemented a PINN hybrid strategy, utilizing Maxwell loss while training to compare against physics-assisted learning strategies. The Maxwell loss evaluates the prediction's degree of satisfaction with Maxwell's equations, thereby providing explicit physical guidance during optimization. Implementation was done by using both MSE and Maxwell loss, with ratio balanced throughout the iterative process (details are provided **in Section S6 of the Supporting Information**).

Model performance was evaluated using the normalized prediction error, defined as follows.

$$L_p(\tilde{y}, y) = \frac{1}{N}\sum_{i=1}^{N}\frac{\|\tilde{y}_i - y_i\|_p}{\|y_i\|_p}$$

Additionally, to assess practical device performance, we extended our evaluation from near-field errors to far-field observables by computing transmission spectra from the predicted fields. Specifically, a near-to-far-field transformation was carried out using Fourier analysis at the output boundary to extract the zeroth-order transmission coefficient. Results were calculated and averaged over 30 layers along the z-direction to ensure stability, and this procedure was applied consistently to both the ground-truth and predicted fields [23]. This enables direct comparison of spectral features, particularly sharp resonant dips, providing a more application-relevant validation of DIRTL for nanophotonic device performance. Evaluation for transmission spectrum accuracy was done via $L_2$ error compared to the true spectrum over 1000 test samples generated with 0.5nm wavelength interval.

## 3. Result and Discussion

### 3.1 Performance Comparison

**Table 1** Training results for baseline and DIRTL FNO, UNet models. Hyperparameter settings for both UNet and FNO are provided in **Section S2 of the Supporting Information**. The training dataset distribution follows the description in **Section 2.3**.

| Model | Training Method | Normalized $L_2$(%) | Transmission Prediction Error (%) |
| --- | --- | --- | --- |
| FNO | Baseline | 5.72 | 2.49 |
| | Baseline + PINN | 5.61 | 2.56 |
| | l-DIRTL | 5.78 | 2.48 |
| | **f-DIRTL** | **2.81** | **1.28** |
| UNet | Baseline | 9.97 | 5.40 |
| | Baseline + PINN | 10.1 | 5.40 |
| | **l-DIRTL** | 9.16 | **4.85** |
| | **f-DIRTL** | **8.59** | 5.43 |

Comparing the f-DIRTL FNO with the baseline FNO reveals a pronounced improvement in prediction accuracy, with the normalized $L_2$ test error decreasing from 5.72% to 2.81%. As shown in **Fig. 4(a)**, f-DIRTL consistently enhances accuracy across both resonant and non-resonant samples. Furthermore, only a very small subset of cases (approximately 0.4%) exhibits slightly degraded performance, further demonstrating the robustness and reliability of the f-DIRTL FNO (see **Section S2 of the Supporting Information**). It is worth noting that we employed such a two-step process due to it being most effective, yielding higher performance compared to multi-step processes. Although it is counterintuitive, multi-step processes fail due to an increase in error during pretrain phases (see **Section S7 of the Supporting Information**).

From the transmission spectrum results in **Table 1** and **Fig. 4(b)**, we observe that for the FNO model, f-DIRTL consistently outperforms the baseline, demonstrating that DIRTL improves not only near-field predictions but also far-field spectral accuracy, which is critical for downstream inverse-design applications. In particular, **Fig. 4(b)** shows that f-DIRTL more accurately predicts the resonant transmission features near 710 nm and 750 nm, which are the spectral regions of greatest interest. This improvement originates from the enhanced field prediction capability of f-DIRTL, where resonant field amplitudes are captured more accurately while overfitting to non-resonant regions is effectively suppressed. Representative field predictions for the given test sample in **Fig. 4(c–g)** illustrate that f-DIRTL yields lower errors than the baseline across multiple wavelengths, improving performance in both resonant (710 nm) and non-resonant cases.

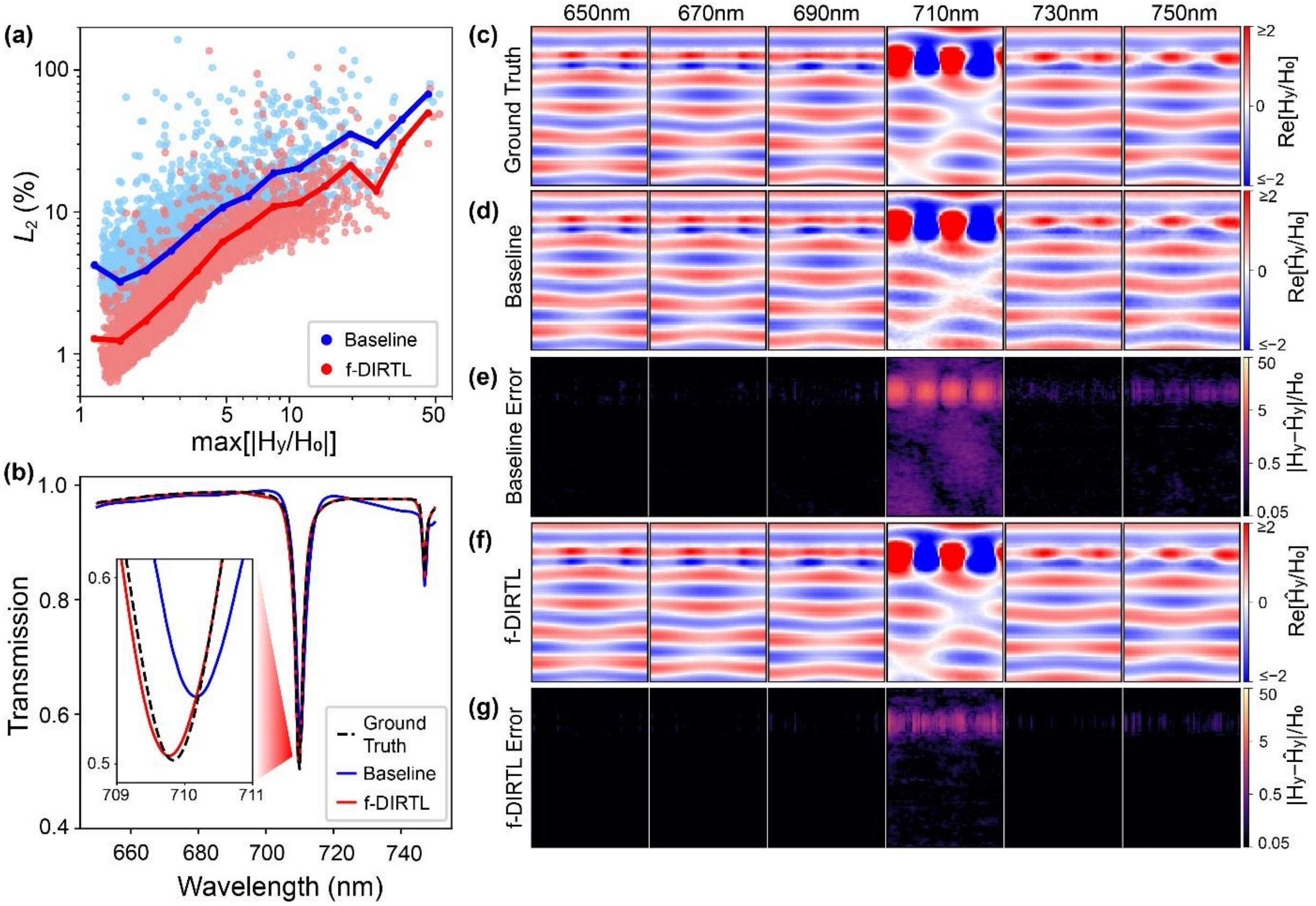

**Figure 4 Comparison test between baseline FNO and f-DIRTL FNO.** Because of the unnoticeable difference between l-DIRTL FNO and baseline FNO, we do not plot the results of l-DIRTL FNO. (a) Error distribution comparison plot between baseline FNO and f-DIRTL FNO. Each light-blue (baseline FNO) and light-coral (f-DIRTL FNO) point represents the prediction error of an individual test sample plotted against its maximum field amplitude. Blue and red markers show the corresponding average errors within logarithmically spaced amplitude bins for the baseline and f-DIRTL models, respectively. The figure is plotted in a log-log scale. Linear scale representation is shown in **Section S2 of the Supporting Information**. (b) Transmission prediction comparison between f-DIRTL FNO (red) and baseline FNO (blue). Ground truth is illustrated with dashed black lines. The inset zooms into the resonant dip near 710nm. (c–g) Field distributions and prediction errors for the test case in (b), at a 20 nm wavelength interval. (c) Ground-truth field distribution. (d) Baseline FNO prediction. (f) f-DIRTL FNO prediction. (e,g) Corresponding prediction-error maps for (e) the baseline FNO and (g) the f-DIRTL FNO. We see that resonance predictions are improved in f-DIRTL FNO.

The benefits of DIRTL are also apparent when trained with UNet architecture. Despite UNet's more limited representational capacity compared to FNO, f-DIRTL UNet still achieves a 1.4% reduction in normalized $L_2$ error relative to its baseline counterpart. The overall inference capability of FNO is superior to UNet for this problem set, due to the ability of FNO to capture global dependencies via operator-level mappings with high parameter efficiency [29, 30]. Interestingly, l-DIRTL UNet also delivers measurable improvements, a trend not observed with FNO. We speculate that this behavior can be attributed to architectural differences. UNet exhibits an explicit

encoder–decoder structure, which enables pretrained representations to be preserved in the encoder while fine-tuning primarily adapts the decoder. In contrast, FNO lacks such structural separation, where all layers jointly contribute to the learned operator. Due to this, fine-tuning induces global parameter updates across the network, making pretrained information less stably preserved during optimization (see **Section S8 of the Supporting Information**).

As for transmission prediction, f-DIRTL UNet exhibits slightly worse performance than the baseline UNet when evaluated using $L_2$ error. However, this discrepancy is primarily driven by improved baseline performance in non-resonant frequency regions, where the model tends to avoid capturing resonant dips altogether. In contrast, f-DIRTL more accurately predicts the position and depth of resonant features, which is better reflected in alternative evaluation metrics such as cosine similarity or $L_4$ error (see **Section S2 of the Supporting Information**).

For Maxwell loss implemented cases, we see little improvement (~0.1% relative $L_2$) in FNO and no improvement in UNet. This is noted by baseline + PINN in **Table 1**. Although the Maxwell loss provides some assistance during FNO training, our proposed two-step strategy delivers substantially greater improvements in predictive accuracy. This contrast suggests that the difficulty in modeling resonant responses cannot be attributed solely to insufficient physical constraints. Rather, it highlights the importance of addressing their inherent multi-scale characteristics and their presence as rare, high-amplitude outliers within the data distribution. By relaxing the resonant features during pretraining, the multi-scale complexity and statistical outlier characteristics are effectively reduced, allowing the model to first learn the global modal structure in a smoother regime. Building on this foundation, the fine-tuning stage then enables the model to accurately capture the high-frequency, localized resonant features.

Additionally, comparisons with alternative training strategies, including oversampling and standard curriculum learning, show inferior performance compared to DIRTL. These results are likely attributed to the lack of discriminative features in the input representation that distinguish between different output behaviors (resonant and non-resonant). Because of this, dataset manipulation strategies such as oversampling become redundant, as they do not introduce meaningful separability in the input space (further analyzed in **Section S5 of the Supporting Information)** [43].

### 3.2 Data Efficiency Analysis

The most notable achievement of this approach lies in its ability to achieve substantial performance gains without increasing the total size of the training dataset or changing the model architecture. Even with a smaller quantity of target-task data, the high-accuracy pretrained core effectively steers the model toward the correct solution, enabling more efficient adaptation than relying on target-task data alone. This behavior differs from standard transfer learning examples, which are typically characterized by large-scale pretraining followed by limited fine-tuning [10,44].

To directly assess the data-efficiency advantages of f-DIRTL FNO over the baseline FNO, we trained both models using varying amounts of training data. Specifically, we prepared datasets consisting of 500, 1,000, 3,000, and 9,000 unique structures, each sampled at 11 wavelength points. For f-DIRTL FNO, half of the samples were used for the lossy pretraining stage and the remaining half for fine-tuning. All training runs used identical learning-rate schedulers as in Section 3.1 to ensure fair comparison.

The resulting prediction errors and distributions are summarized in **Fig. 5(a)**. Across all dataset sizes, f-DIRTL FNO consistently outperforms the baseline FNO. Notably, when trained with 9,000 structures, the baseline FNO achieves a normalized $L_2$ error of 5.72%, which is comparable to the f-DIRTL FNO trained with only 1,000 structures (6.25%). This quantitative comparison suggests that f-DIRTL can achieve similar accuracy with nearly an order of magnitude fewer training samples, corresponding to an effective ~9 times improvement in data efficiency. Such a gain significantly enhances the practicality and productivity of neural operator-based EM surrogate solvers.

### 3.3 Robustness Assessment for f-DIRTL

To evaluate the robustness of f-DIRTL FNO to different training conditions, we conducted two hyperparameter sweeps. First, we varied the number of Fourier layers used during pretraining, transferring and freezing 1 to 10 layers into a 10-layer fine-tuned FNO model. Second, we investigated the effect of the pretrain-to-fine-tune data ratio, where a ratio of 0.1 indicates that the pretraining dataset consists of 10% of the total dataset, with the remaining 90% being for fine-tuning. Ratios from 0.1 to 0.9 were tested in increments of 0.1. All experiments used a total of 3,000 randomly sampled structures with wavelengths from 650 nm to 750 nm at 10 nm intervals (3,000 samples $\times$ 11 wavelength points, in total 33,000 training samples). These sweeps allowed us to assess how f-DIRTL performance depends on both dataset distribution and model design.

In the second hyperparameter sweep, we varied the number of Fourier layers used during pretraining and the number of layers transferred to the fine-tuned model. This enabled us to evaluate how different transfer strategies affect the final prediction accuracy. Adjusting the number of transferred layers also provided insight into how information learned during pretraining is distributed throughout the model. For this study, both the pretraining and transfer stages used 1,500 randomly sampled structures (1,500 samples $\times$ 11 wavelength points, 16,500 training samples for each phase, total 33,000 training samples). We tested pretrained models with 1 to 10 Fourier layers and transferred anywhere from a single layer to the entire sequence.

The results of these sweeps are shown in **Fig. 5(b,c)**. In **Fig. 5(b)**, the prediction error increases sharply when pretraining more than eight layers. This is primarily due to underfitting during the fine-tuning stage, as there are too few trainable parameters remaining. In contrast, pretraining and transferring fewer than eight layers results in a model that is robust to dataset distribution, with prediction errors around or below 4%, with the best case reaching 3.6% (baseline FNO error for 3k random structures is 7.9%). This behavior reflects a simple trade-off: using more data for pretraining improves the accuracy of the pretrained core, while having sufficient data in fine-tuning ensures accurate adaptation. This test suggests that, when applying this strategy, the dataset can be prepared without excessive concern for data balance, provided that the pretrain and fine-tune dataset sizes are roughly comparable. As for the number of pretrained layers, we find that pretraining and transferring 4 to 7 layers yields the best performance. Using fewer layers leads to underfitting during the pretraining stage, while using more layers limits the number of trainable parameters during fine-tuning, resulting in severe underfitting in the transfer stage.

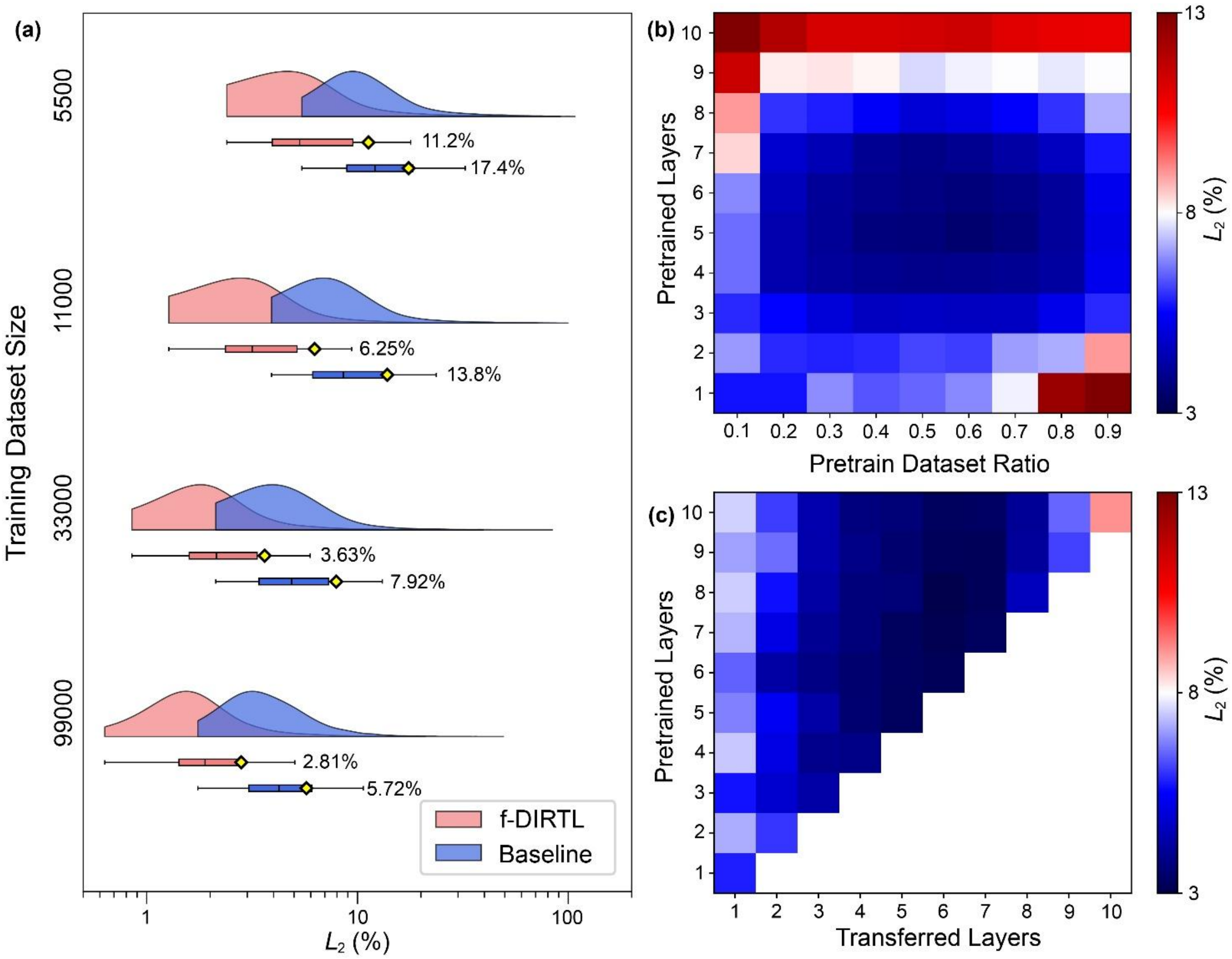

**Figure 5 Robustness test on training dataset size and training condition** (a) Density plot of normalized $L_2$ errors with varying training set sizes. Density plots show kernel density estimates of the normalized $L_2$ error distributions for both models across different training-set sizes. Beneath each density curve, horizontal box plots summarize the corresponding distributions, showing the median (notch), quartiles, and whiskers, with yellow diamonds marking the mean values annotated alongside each box. (b) Hyperparameter sweep altering ratio of pretrain dataset and number of Fourier layers used in pretrain stage. (c) Hyperparameter sweep altering number of Fourier layers in pretrain stage and amount of layers transferred in the fine-tune stage. For both (b) and (c), the colormap indicates that shades of blue are better than baseline, and shades or red are worse than baseline.

In **Fig. 5(c)**, we observe that, regardless of the total number of pretrained layers, transferring between four and seven layers yields the best results. Similar to the case above, for larger pretrained models, partially transferring the network allows the fine-tuned model to retain sufficient trainable parameters for adapting to the target problem. Conversely, for smaller pretrained models, transferring nearly all pretrained layers produced the best performance, suggesting that most of the learned physical information is in the deeper Fourier layers. Nevertheless, even transferring just one or two layers improves fine-tuned accuracy beyond the baseline, indicating that essential physical features are also embedded within the lower layers of the network.

Meanwhile, both the f-DIRTL FNO and the baseline FNO exhibit excellent interpolation performance at unseen wavelengths, demonstrating robust inference capability. This inference ability is further enhanced during the lossy pretraining phase, where the pretrained model demonstrates interpolation over broader wavelength ranges as well as wavelength extrapolation. These results are directly utilized as multi-tasking of the DIRTL framework, in which a coarsely trained pretrained core is effectively adapted to more specific target tasks during the fine-tuning phase. Details on this are demonstrated in **Section S9 of the Supporting Information**.

## 4. Conclusion

In this work, we introduced dissipative relaxation transfer learning (DIRTL), a physics-inspired training strategy designed to resolve a central challenge in electromagnetic surrogate modeling: accurately learning high-amplitude resonant responses that appear as statistical outliers within predominantly non-resonant datasets. Drawing an analogy to loss-regularized optimization in photonic design, DIRTL employs a two-stage curriculum in which resonant features are first attenuated using fictitious material absorption and are later reintroduced during fine-tuning. This process enables the model to first acquire a smooth, global modal structure before confronting the sharp, nonconvex characteristics of high-Q responses.

Our experimental results show that DIRTL significantly improves surrogate-solver accuracy, while preserving the total amount of training data required, for the given problem dataset. In particular, the f-DIRTL variant, where pretrained layers are transferred and frozen within a deeper fine-tuning model, achieves more than a twofold reduction in prediction error compared with baseline FNO training. In addition, our dataset-scaling experiments show that the f-DIRTL FNO achieves comparable prediction accuracy using only about one-ninth of the total training data required by the baseline model. Such results are uncommon in conventional transfer learning, underscoring the pretrained core's ability to provide physically meaningful guidance. Overall, the accurately pretrained core effectively transfers knowledge of global modal behavior to the fine-tuned model, enabling concrete adaptation to the target lossless system with highly resonant features (for visualization of the learning dynamics, see **Section S10 of the Supporting Information**).

Comprehensive parameter sweeps further show that DIRTL retains strong performance across diverse dataset splits and transfer configurations, highlighting both its robustness and its ease of integration into existing workflows. In particular, experiments varying the number of transferred Fourier layers demonstrate that most of the physically meaningful representations are preserved when the entire sequence of pretrained layers is transferred. Nevertheless, even transferring only one or two layers yields improvements over the baseline, suggesting that essential physical features are also encoded in the shallow layers. Across architectures, consistent gains are observed for both FNO and UNet, suggesting compatibility with multiple surrogate architectures.

While the preceding sections focus on the 1D binary grating as a controlled benchmark, an important next step is to evaluate how the DIRTL framework extends to more realistic electromagnetic problems. In practical nanophotonic applications, the governing physics becomes significantly more complex, involving lossy materials (see **Section S11 of the Supporting Information**), multiple interacting resonances, and various nonlocal effects. These factors increase the difficulty of accurate field prediction, not only for resonant responses but also for capturing the global field structure.

To address this, we further demonstrate DIRTL on a two-dimensional multilayer device, where both local and nonlocal modal interactions give rise to a substantially more complex electromagnetic landscape. In this setting, we observe improvements in prediction accuracy for f-DIRTL compared to the baseline (detailed in **Section S12 of the Supporting Information**). Meanwhile, the performance gain of DIRTL is reduced for the more complex problem compared to the 1D binary grating case, due to the increased difficulty arising from multiple modal interactions. In such regimes, the learning task extends beyond capturing localized resonant features to accurately modeling the global field distribution shaped by the interplay of multiple modes. This challenge is further exacerbated when training data is limited, leading to incomplete learning of the underlying global structure.

As a result, the effectiveness of the current DIRTL framework in highly complex systems is constrained by both data availability and model capacity. This suggests that scaling DIRTL to more demanding problems, such as full 3D field simulations with 2D periodic structures, will require sufficiently large and representative datasets, as well as models capable of capturing increasingly rich multiscale interactions. In conclusion, achieving robust surrogate solvers in these regimes cannot rely on improvements in a single component alone. Instead, future progress will require the combined advancement of dataset construction, model architecture design, physics-informed learning approaches, and transfer learning strategies. A holistic integration of these elements is essential for enabling accurate and scalable surrogate modeling in challenging nanophotonic inverse-design problems.

Building on these considerations, the present results position DIRTL as a data-efficient and physically grounded training framework that bridges photonic optimization principles with transfer learning. By structuring the training process to first learn smooth global modal behavior and subsequently refine high-Q resonant features, DIRTL enables neural networks to more effectively capture the underlying physics with limited data. Given its model-agnostic nature and demonstrated robustness, the framework provides a practical pathway toward more reliable surrogate modeling, with strong potential for integration into modern inverse-design workflows.

## Acknowledgements

This work was supported by the Culture, Sports and Tourism R&D Program through the Korea Creative Content Agency (KOCCA), funded by the Ministry of Culture, Sports and Tourism (RS-2024-00332210), and by the National Research Foundation of Korea (NRF) grants funded by the Ministry of Science and ICT (RS-2024-00414119 and RS-2024-00416583) and the Ministry of Education (RS-2024-00360139), Republic of Korea.

## Author Contributions

S.N., C.Y.P., and M.S.J. conceived the idea and designed the methodology. S.N., C.Y.P., and M.S.J. conducted theoretical analysis. S.N. performed simulations, model training, and organization of data. S.N., C.Y.P., and M.S.J. analyzed the data and wrote the manuscript. C.Y.P. and M.S.J. supervised the project.

## Competing Interests

The authors declare no competing interests.

## Data and Materials Availability

All data needed to evaluate the conclusions in this study are presented in the paper and in the Supporting Information. We open our source code at a freely accessible repository:

https://github.com/jLabKAIST/dissipative-relaxation-transfer-learning

## Supporting Information

Details of the RCWA convergence tests, network architectures and training hyperparameters for FNO and UNet models, quasinormal-mode analysis, fictitious loss tuning, additional baseline training comparison, implementation of the Maxwell loss, multi-step DIRTL analysis, model comparisons for l-DIRTL, multi-task wavelength interpolation and extrapolation results, f-DIRTL internal visualizations, DIRTL implementation on lossy target data, and generalization to 2D multilayer structures are included.

# Supporting Information

## Data-Efficient Electromagnetic Surrogate Solver Through Dissipative Relaxation Transfer Learning

Sunghyun Nam[1], Chan Y. Park[2*], and Min Seok Jang[1*]

[1]School of Electrical Engineering, Korea Advanced Institute of Science and Technology, Daejeon 34141, Republic of Korea

[2]KC Machine Learning Lab, Seoul 06181, Republic of Korea

*Corresponding author.

Email: Sunghyun Nam: felix0616@kaist.ac.kr, Chan Y. Park: chan.y.park@kc-ml2.com,
Min Seok Jang: jang.minseok@kaist.ac.kr

## S1. RCWA Engine Convergence Test

To generate the training and test datasets used for neural operator training, we employed the meent rigorous coupled-wave analysis (RCWA) engine [S1]. The selection of the Fourier order was determined through a convergence study conducted on a highly resonant test structure. We first generated 1,000 random structural samples at 700 nm using a high reference setting of 101 Fourier orders, and identified a structure exhibiting strong resonance. Using this structure, we evaluated the field convergence across varying Fourier orders. Convergence was evaluated by comparing the results obtained at each Fourier order with those computed using a very high reference setting of 1281 Fourier orders. Error rate was measured using normalized $L_2$ error.

From this analysis, 81 Fourier orders were found to provide a suitable balance between numerical accuracy and computational efficiency for large-scale dataset generation, retaining below 1% normalized $L_2$ error. The simulation results are shown in **Fig. S1(a)**, while the error for each Fourier order is shown in **Fig. S1(b)**.

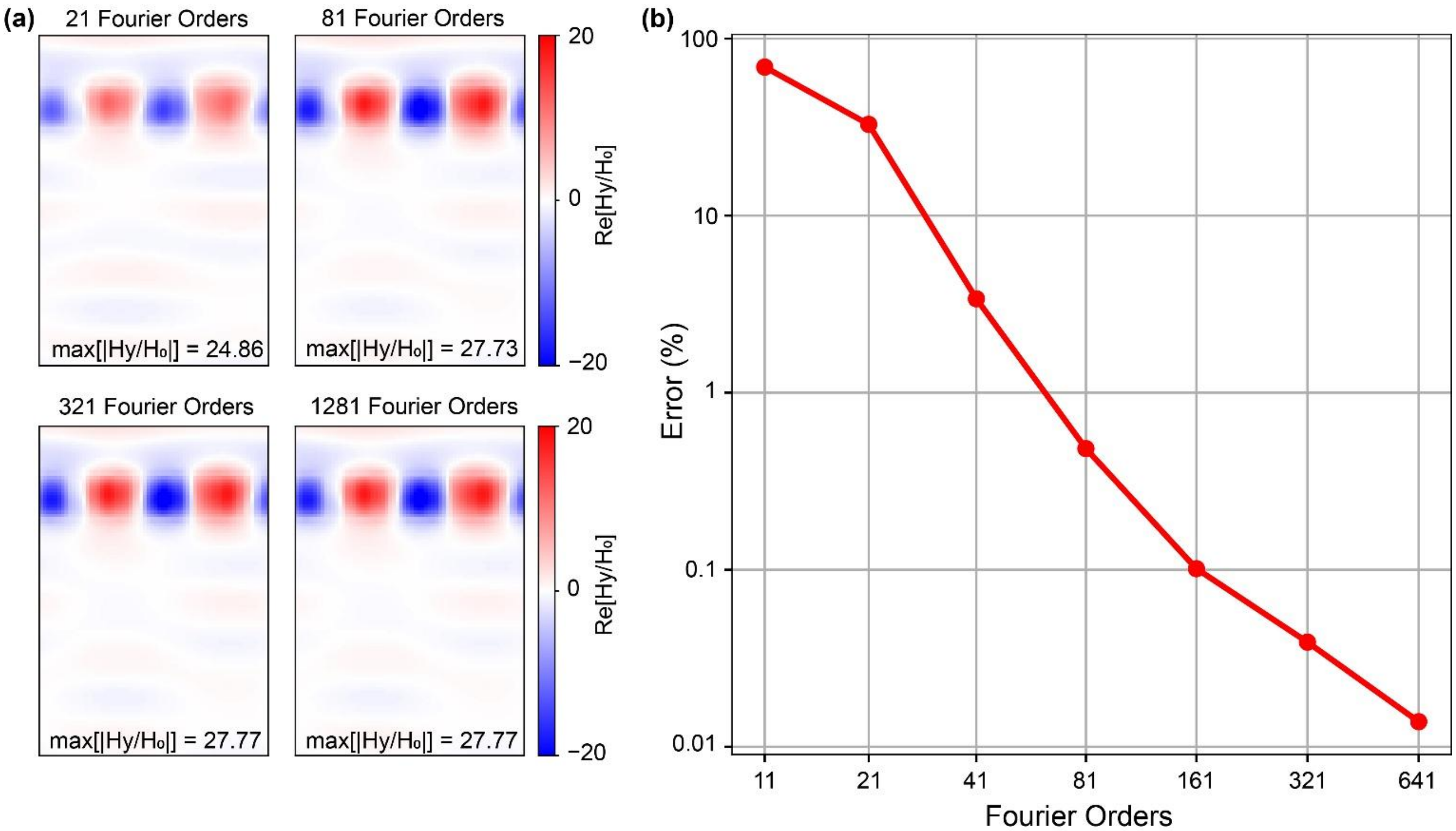

**Figure S1 Convergence test for RCWA engine under various Fourier orders.** Bottom right result shows that for FEM solver.

## S2. Training Details on FNO and UNet

The hyperparameters for Fourier Neural Operator (FNO) and UNet used in this work are shown in **Table S1**, while the learning rate scheduler parameters are shown in **Table S2**. All training and test of models were done on NVIDIA® GeForce RTX™ 3080 GPU. The detailed architecture and l-DIRTL, f-DIRTL process for UNet is shown in **Fig S2** [S2]. The training learning curve is shown in **Fig. S3**. **Figure S4** shows overfitting examples, where neural network (NN) models falsely predict resonances where only background scattering occurs. **Figure S4** also demonstrates how f-DIRTL FNO suppresses such cases, increasing overall accuracy.

**Figure S5** represents the error scatter plot comparison between f-DIRTL and baseline FNO. In **Fig. S5(a, b)**, we can observe that f-DIRTL FNO outperforms the baseline for both resonant and non-resonant test samples. **Figure S5(c, d)** illustrates that on a per-sample basis, the majority of predictions are more precise than those of the baseline model. Only a very small subset of cases (about 0.4%) exhibits slightly degraded performance, corresponding to highly resonant geometries where both models struggle. This suggests that although f-DIRTL effectively guides the model toward the correct solution manifold, certain resonant cases remain difficult and may warrant further analysis.

For the UNet model, the baseline achieves slightly better performance in terms of the $L_2$ error metric for transmission prediction. However, this is primarily driven by improved accuracy in non-resonant regions, while the model tends to avoid capturing resonant dips altogether. In contrast, f-DIRTL UNet more effectively predicts the overall spectral shape and the positions of resonant features, as shown in **Fig. S6**. Although both models exhibit noticeable errors, f-DIRTL better reproduces resonance-associated spectral features despite the slightly worse overall $L_2$ metric. This observation is further supported by alternative metrics that emphasize resonant behavior, such as $L_4$ error (14.2% for baseline, 12.3% for f-DIRTL) and cosine similarity (0.9985 for baseline, 0.9989 for f-DIRTL), where f-DIRTL demonstrates superior performance [S3].

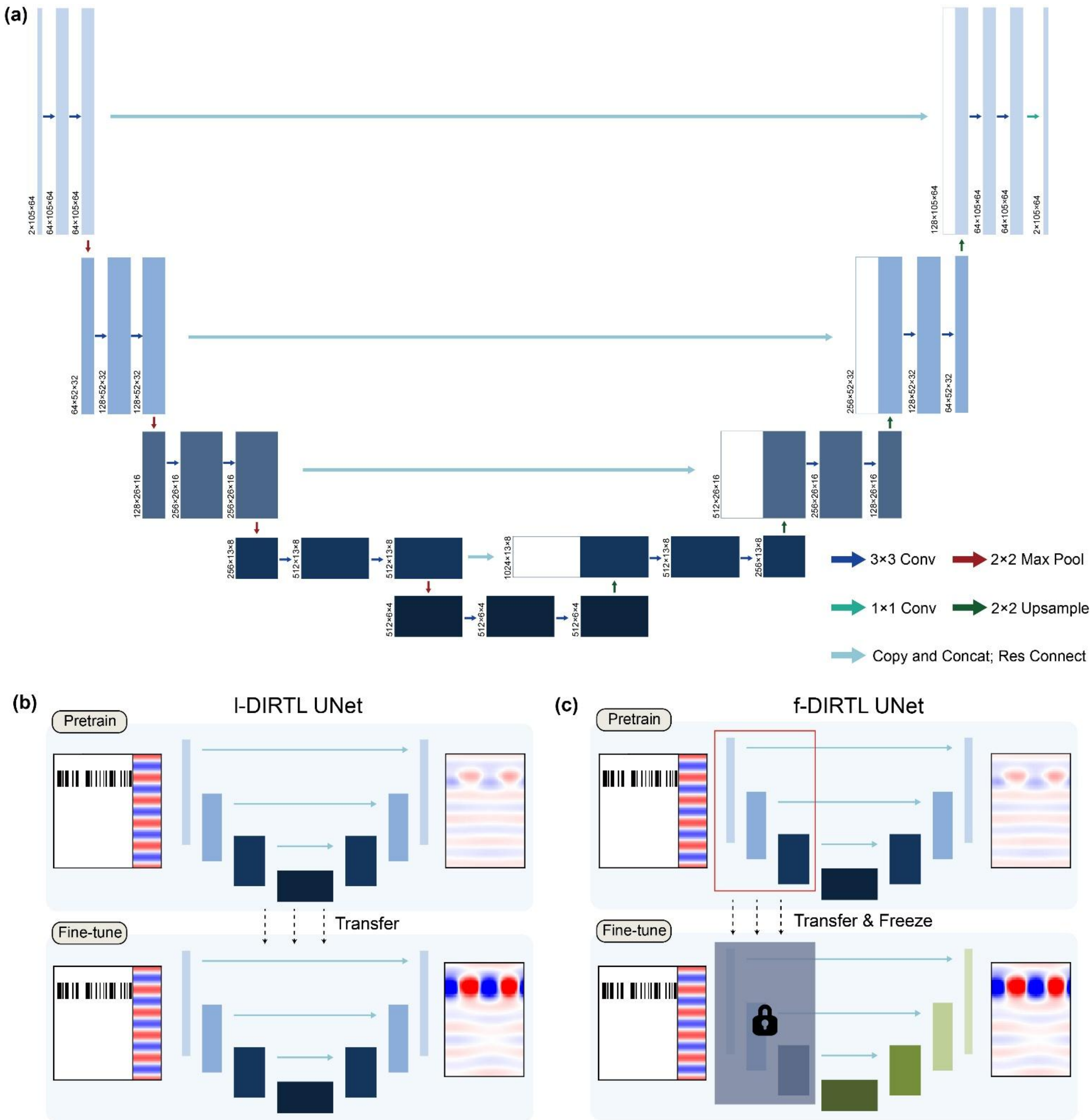

**Figure S2 Architecture and DIRTL process for UNet.** (a) Detailed architecture for UNet used in EM inference for this work [S2]. (b) l-DIRTL and (c) f-DIRTL process for UNet. In f-DIRTL, we transfer and freeze the downsampling encoding phase inside UNet architecture.

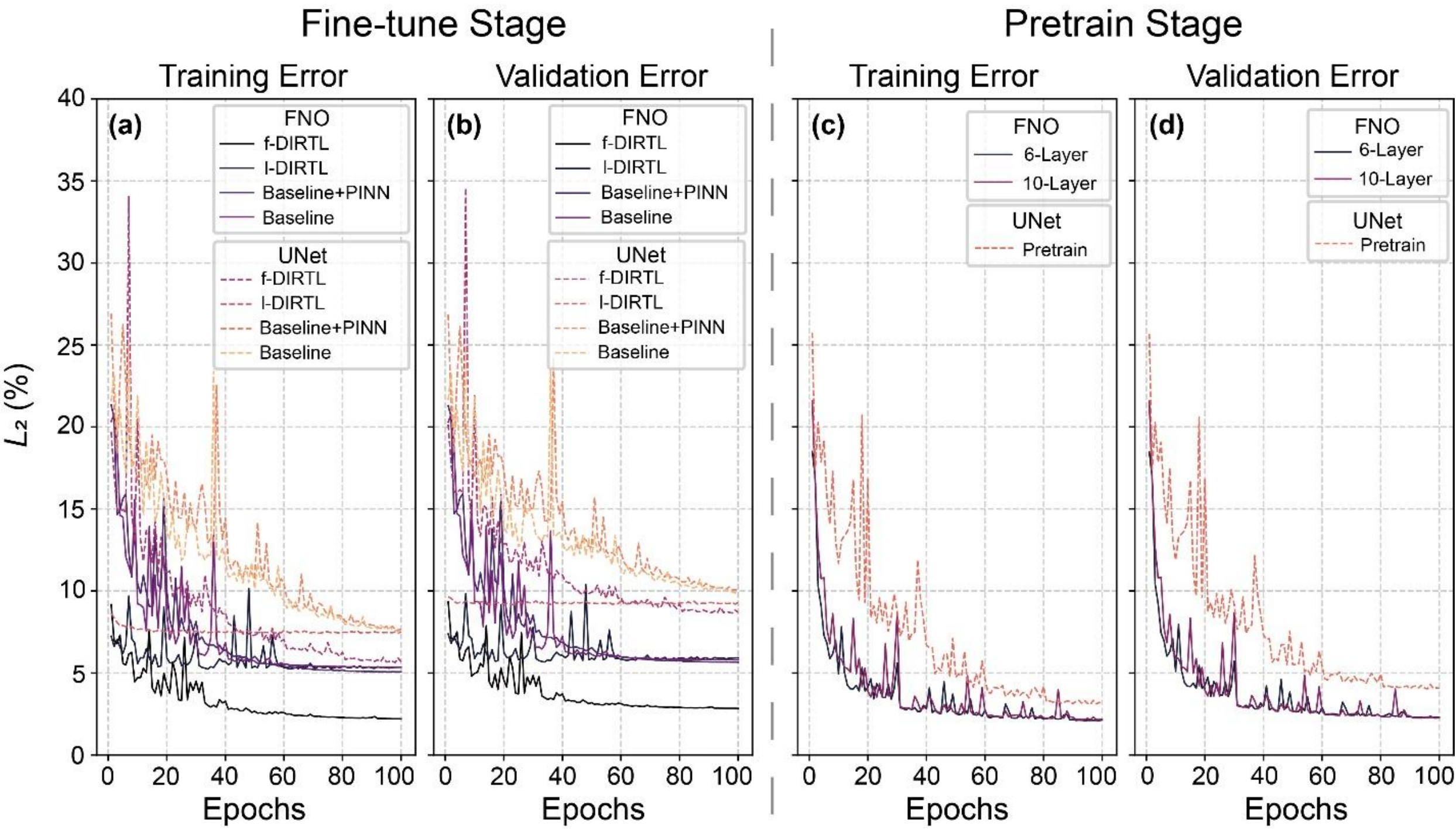

**Figure S3 Learning curves of the models trained for comparison test.** Solid lines are those for FNO, while dotted lines are those for UNet. (a, b) are training error and validation error for the fine-tune stages and the baseline training, while (c, d) show training and validation for the pretrain stage.

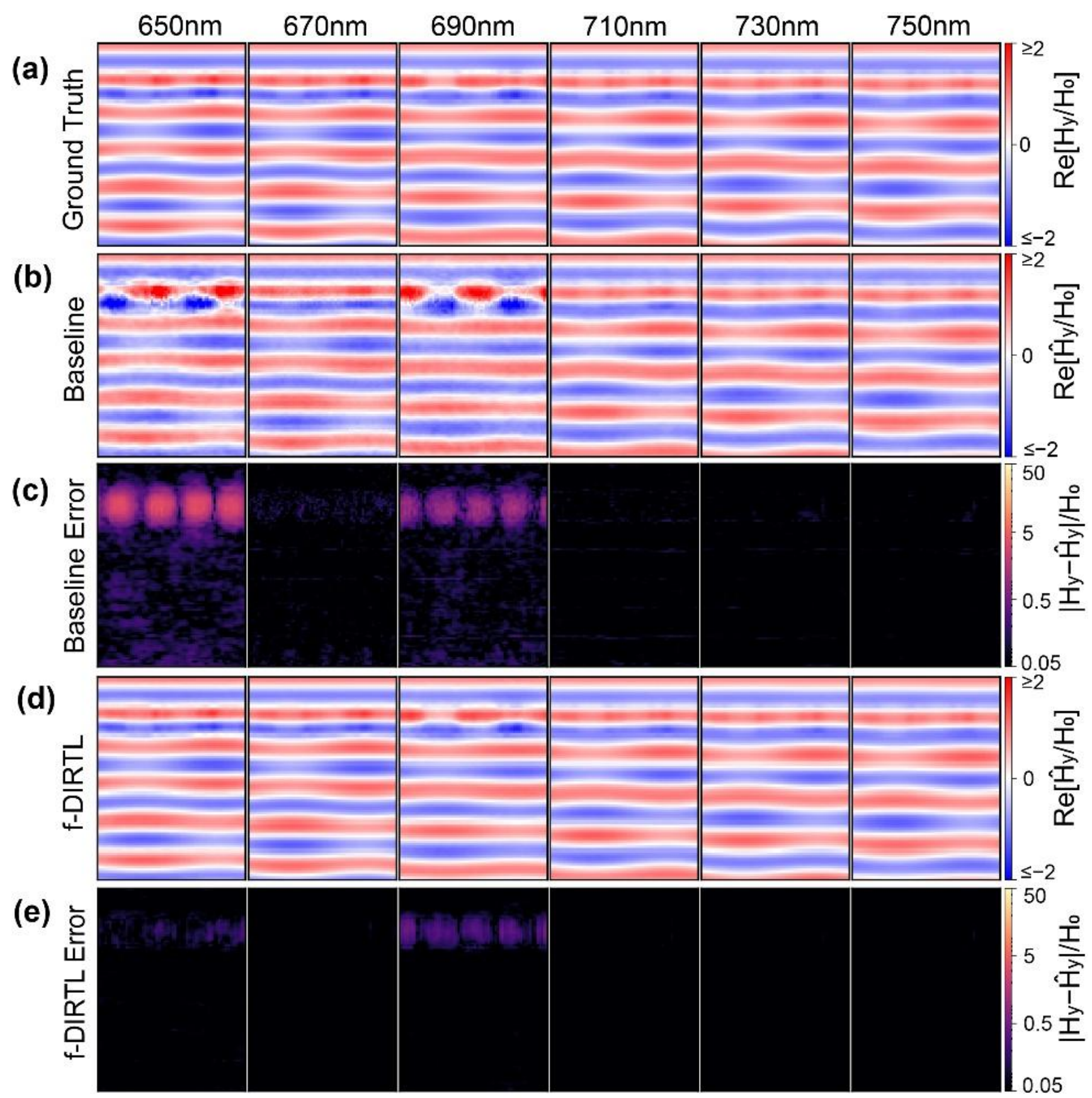

**Figure S4 Overfitting problem of baseline FNO.** Field distributions and prediction errors for a randomly selected test case at a 20 nm wavelength interval. (a) Ground-truth field distribution. (b) Baseline FNO prediction. (d) f-DIRTL FNO prediction. (c,e) Corresponding prediction-error maps for (c) the baseline FNO and (e) the f-DIRTL FNO. We see falsely predicted resonances in baseline FNO prediction, which doesn't happen for f-DIRTL FNO prediction.

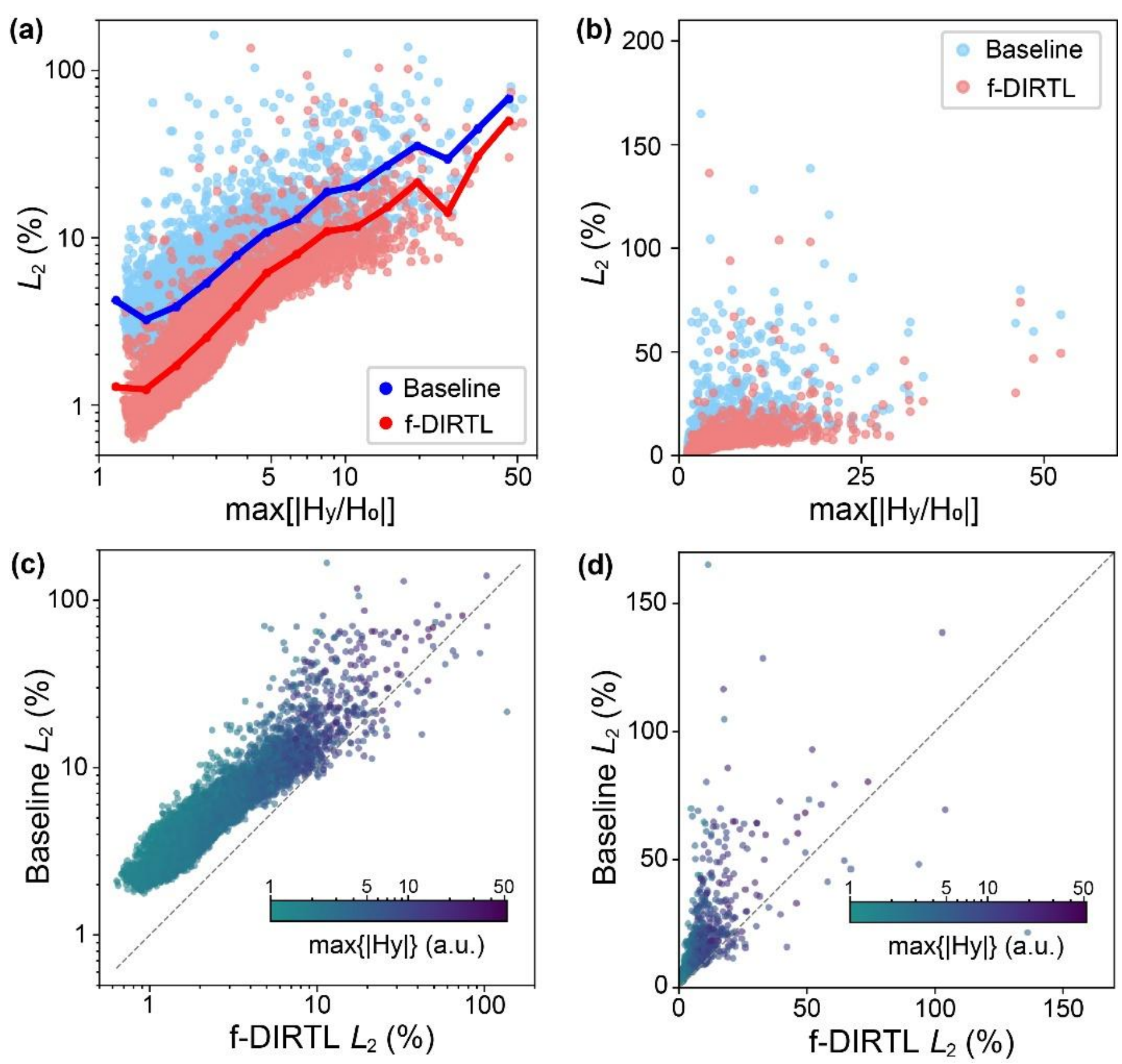

**Figure S5 Error scatter plot comparison between f-DIRTL and baseline FNO.** (a) Error distribution comparison plot between baseline FNO and f-DIRTL FNO. Each light-blue (baseline FNO) and light-coral (f-DIRTL FNO) point represents the prediction error of an individual test sample plotted against its maximum field amplitude. Blue and red markers show the corresponding average errors within logarithmically spaced amplitude bins for the baseline and f-DIRTL models, respectively. The figure is plotted in a log-log scale. (b) Error distribution comparison plot between baseline FNO and f-DIRTL FNO in a linear-linear scale. (c) Per-sample error comparison. Each point compares the baseline and f-DIRTL prediction errors for the same sample. Relative to the diagonal y=x, the strong bias of points toward the baseline-error side indicates that f-DIRTL achieves consistently lower errors across the vast majority of test cases. The figure is plotted in a log-log scale. (d) Per-sample error comparison plotted in a linear-linear scale.

**Table S1** FNO and UNet hyperparameters

| Model | Total Parameters | Hyperparameters | | | |
|---|---|---|---|---|---|
| FNO | 10,496,162 | Modes | 16 | Layers | 10 |
| | | Width | 32 | Activation | GeLU |
| UNet | 17,262,466 | Depth | 4 | Initial Feature Maps | 64 |
| | | Convolution Kernel | 3 × 3 | Pooling | Maxpool |
| | | Upsampling Kernel | 2 × 2 | Maxpool Kernel | 2 ×2 |
| | | Scaling Factor | 2 | Activation | ReLU |

**Table S2** Learning rate scheduler parameter for each training process

| Model | Training Phase | Learning Rate Parameters | | |
|---|---|---|---|---|
| | | Initial Lr | Step Size | Gamma |
| FNO | Baseline | 5e-4 | 10 | 0.5 |
| | 6-Layer Pretrain | 1e-3 | 30 | 0.5 |
| | 10-Layer Pretrain | 5e-4 | 30 | 0.5 |
| | f-DIRTL | 5e-3 | 10 | 0.7 |
| | l-DIRTL | 5e-5 | 30 | 0.5 |
| UNet | Baseline | 1e-4 | 20 | 0.5 |
| | Pretrain | 2e-4 | 20 | 0.5 |
| | f-DIRTL | 7e-5 | 20 | 0.6 |
| | l-DIRTL | 1e-6 | 15 | 0.3 |

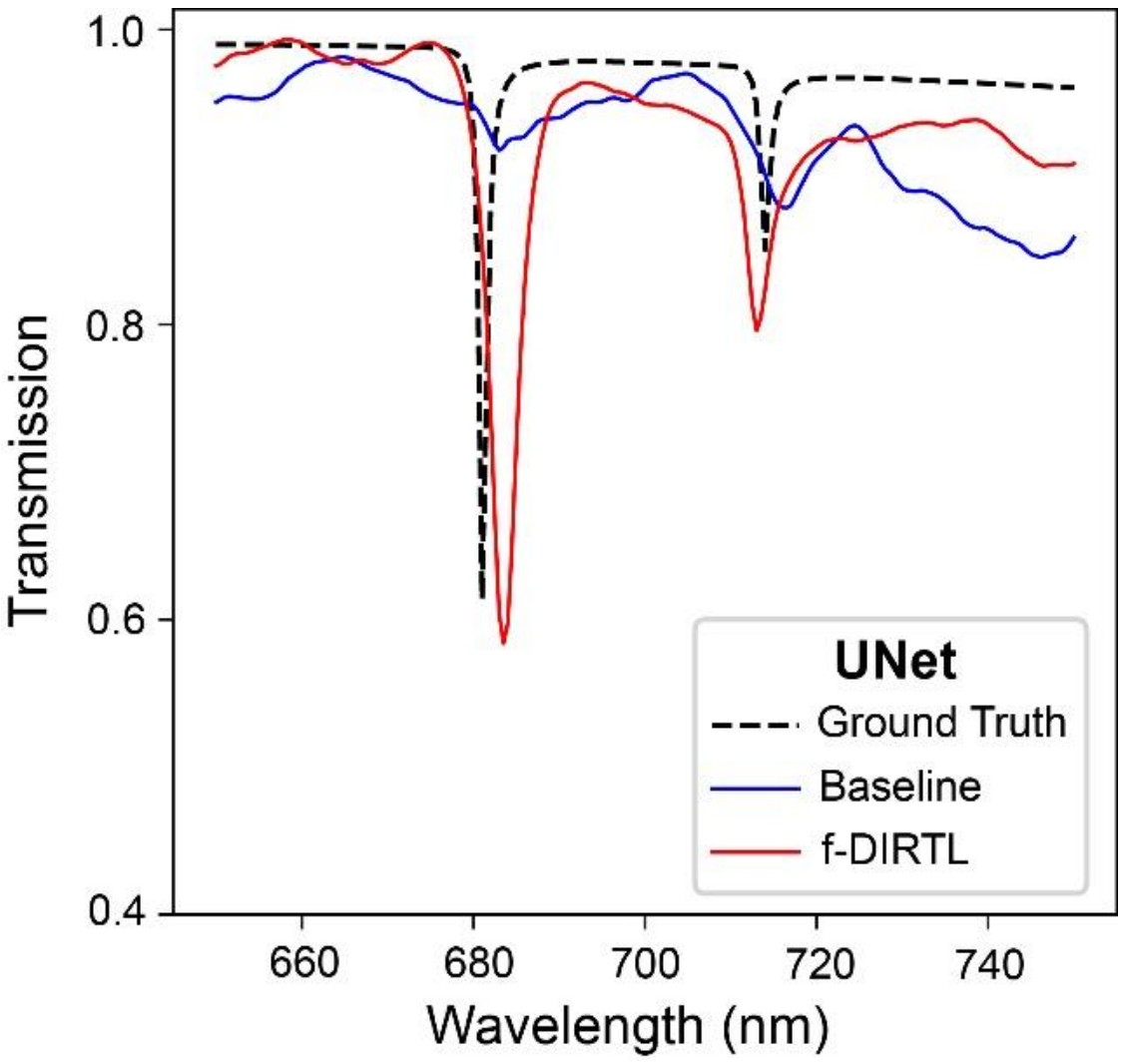

**Figure S6 Transmission spectrum prediction comparison for baseline and f-DIRTL UNet.** Although f-DIRTL UNet shows worse prediction $L_2$ error-wise, it predicts the overall response picture better than the baseline in terms of where spectral dips appear.

### S3. Quasinormal Mode Decomposition of Field Response

The resonant modes of an electromagnetic (EM) system can be obtained by solving a sourceless Maxwell wave equation, which forms an eigenvalue problem for the electric field. These eigenmodes are characterized by complex frequency eigenvalues, known as resonant frequencies or quasinormal modes (QNMs) [S4]. Formally, for systems with free-space magnetic permeability $\mu_0$, the EM problem can be expressed as

$$\frac{1}{\mu_0}\nabla\times\nabla\times\mathbf{E}-\omega^2\varepsilon(\mathbf{r})\mathbf{E}=i\omega\mathbf{J}(\mathbf{r})$$

The corresponding sourceless eigenproblem defining QNMs is then given as

$$\nabla\times\nabla\times\tilde{\mathbf{E}}_m=\left(\frac{\tilde{\omega}_m}{c}\right)^2\epsilon(\mathbf{r})\tilde{E}_m,\qquad \varepsilon(\boldsymbol{r})=\varepsilon_0\epsilon(\mathbf{r})$$

where $\tilde{\mathbf{E}}_m$ and $\tilde{\omega}_m$ denoth the mth QNM and its corresponding complex eigenfrequecy. In practice, these modes can be computed using eigenfrequency solvers or complex-frequency sweeps. For example, using the S4 RCWA solver that supports complex frequency analysis, the reflection spectrum reveals a pole near 700 nm, slightly below the real axis in the complex frequency plane (see **Fig. 2(a)**), corresponding to a resonant QNM [S5].

Now, to study the effect of small material absorption, we introduce a small complex permittivity term $i\delta\epsilon(r)$ to the system, which can be written in a perturbed eigenproblem like below.

$$\nabla\times\nabla\times\left(\tilde{\mathbf{E}}_m+\delta\tilde{\mathbf{E}}_m\right)=\left(\frac{\tilde{\omega}_m+\delta\tilde{\omega}_m}{c}\right)^2\left(\epsilon(\mathbf{r})+i\delta\epsilon(\mathbf{r})\right)\left(\tilde{\mathbf{E}}_m+\delta\tilde{\mathbf{E}}_m\right)$$

Retaining only first-order perturbation terms, we have

$$\nabla\times\nabla\times\delta\tilde{\mathbf{E}}_m=i\left(\frac{\tilde{\omega}_m}{c}\right)^2\delta\epsilon(\mathbf{r})\tilde{\mathbf{E}}_m+\left(\frac{\tilde{\omega}_m}{c}\right)^2\epsilon(\mathbf{r})\delta\tilde{\mathbf{E}}_m+2\left(\frac{\tilde{\omega}_m}{c^2}\right)\delta\tilde{\omega}_m\epsilon(\mathbf{r})\tilde{\mathbf{E}}_m$$

Now, as the new eigenmode $\tilde{\mathbf{E}}_m+\delta\tilde{\mathbf{E}}_m$ can be described as a span of QNMs

$$\tilde{\mathbf{E}}_m+\delta\tilde{\mathbf{E}}_m=\sum_n\alpha_n\tilde{\mathbf{E}}_n$$

we can describe the correction term as span of eigenmodes not including original m[th] eigenmode [S6]. Therefore, ultilizing the orthornormality of QNMs

$$\delta\tilde{\mathbf{E}}_m=\sum_{n\neq m}\alpha_n\tilde{\mathbf{E}}_n\,,\int\tilde{\mathbf{E}}_m^*\cdot\tilde{\mathbf{E}}_n dv=\delta_{mn}$$

Taking inner product for first order equation with $\tilde{\mathbf{E}}_m^*$, we get

$$i\left(\frac{\widetilde{\omega}_m}{c}\right)^2 \int \tilde{\mathbf{E}}_m^* \cdot \delta\epsilon(\mathbf{r})\tilde{\mathbf{E}}_n dv = -2\left(\frac{\widetilde{\omega}_m}{c^2}\right)\delta\widetilde{\omega}_m \int \tilde{\mathbf{E}}_m^* \cdot \epsilon(\mathbf{r})\tilde{\mathbf{E}}_n dv$$

From this, we get the well-known first-order correction for eigenfrequency $\delta\widetilde{\omega}_m$.

$$\delta\widetilde{\omega}_m = -i\frac{\widetilde{\omega}_m \int \tilde{\mathbf{E}}_m^* \cdot \delta\epsilon(\mathbf{r})\tilde{\mathbf{E}}_n dv}{2\int \tilde{\mathbf{E}}_m^* \cdot \epsilon(\mathbf{r})\tilde{\mathbf{E}}_n dv}$$

Hence, the perturbation $i\delta\epsilon(\mathbf{r})$ contributes only to the imaginary part of the eigenfrequency, confirming that small fictitious loss affects resonance linewidth, but not the real resonance position. This effect is well shown in the pole shift in Fig. 2(a).

Now, for the mode correction terms, substituting $\delta\tilde{\mathbf{E}}_m = \sum_{n\neq m}\alpha_n\tilde{\mathbf{E}}_n$ to first order perturbation equation, we obtain

$$\sum_{n\neq m}\alpha_n\nabla\times\nabla\times\tilde{\mathbf{E}}_n = i\left(\frac{\widetilde{\omega}_m}{c}\right)^2\delta\epsilon(r)\tilde{\mathbf{E}}_m + \left(\frac{\widetilde{\omega}_m}{c}\right)^2\epsilon(\mathbf{r})\sum_{n\neq m}\alpha_n\tilde{\mathbf{E}}_n + 2\left(\frac{\widetilde{\omega}_m}{c^2}\right)\delta\widetilde{\omega}_m\epsilon(\mathbf{r})\tilde{\mathbf{E}}_m$$

As $\nabla\times\nabla\times\tilde{\mathbf{E}}_n = \left(\frac{\widetilde{\omega}_n}{c}\right)^2\epsilon(\mathbf{r})\tilde{\mathbf{E}}_n$, we can rewrite as

$$\sum_{n\neq m}\alpha_n\left(\frac{\widetilde{\omega}_n{}^2 - \widetilde{\omega}_m{}^2}{c^2}\right)\epsilon(\mathbf{r})\tilde{\mathbf{E}}_n = i\left(\frac{\widetilde{\omega}_m}{c}\right)^2\delta\epsilon(\mathbf{r})\tilde{\mathbf{E}}_m + 2\left(\frac{\widetilde{\omega}_m}{c^2}\right)\delta\widetilde{\omega}_m\epsilon(\mathbf{r})\tilde{\mathbf{E}}_m$$

Now, taking inner product with $\widetilde{\mathbf{E}_n}^*$, we get

$$\alpha_n\left(\frac{\widetilde{\omega}_n{}^2 - \widetilde{\omega}_m{}^2}{c^2}\right)\epsilon(\boldsymbol{r}) = i\left(\frac{\widetilde{\omega}_m}{c}\right)^2\int \tilde{\mathbf{E}}_m^* \cdot \delta\epsilon(\mathbf{r})\tilde{\mathbf{E}}_m dv$$

$$\alpha_n = \frac{i\widetilde{\omega}_m{}^2\int \tilde{\mathbf{E}}_m^* \cdot \delta\epsilon(\mathbf{r})\tilde{\mathbf{E}}_m dv}{\left(\widetilde{\omega}_n{}^2 - \widetilde{\omega}_m{}^2\right)\epsilon(\mathbf{r})}$$

$$\delta\tilde{\mathbf{E}}_m = i\sum_{n\neq m}\frac{\widetilde{\omega}_m{}^2\int \tilde{\mathbf{E}}_m^* \cdot \delta\epsilon(\mathbf{r})\tilde{\mathbf{E}}_m dv}{\left(\widetilde{\omega}_n{}^2 - \widetilde{\omega}_m{}^2\right)\epsilon(\mathbf{r})}\tilde{\mathbf{E}}_n$$

By this, we can also derive the mode correction term via first order perturbation. Meanwhile, now approaching the Maxwell equation in terms of Green's function like below,

$$\frac{1}{\mu_0}\nabla\times\nabla\times G(\mathbf{r},\mathbf{r}',\omega) - \omega^2\varepsilon(\mathbf{r})G(\mathbf{r},\mathbf{r}',\omega) = \delta(\mathbf{r}-\mathbf{r}')$$

we can decompose the Green's function via QNMs like below [S4]

$$G(\mathbf{r},\mathbf{r}',\omega) = -\sum_m \frac{\tilde{\mathbf{E}}_m(\mathbf{r}) \otimes \tilde{\mathbf{E}}_m(\mathbf{r}')}{\omega(\omega - \tilde{\omega}_m)}$$

which yields the field response

$$\mathbf{E}(\mathbf{r}) = \int G(\mathbf{r},\mathbf{r}',\omega) \cdot i\omega \mathbf{J}(\mathbf{r}) dv' = -i \sum_m \frac{\tilde{\mathbf{E}}_m(\mathbf{r}) \otimes \tilde{\mathbf{E}}_m(\mathbf{r}')}{(\omega - \tilde{\omega}_m)} \cdot \mathbf{J}(\mathbf{r}) dv'$$

For non-resonant structures, the QNM poles are located far from the real frequency axis. Consequently, the term $\omega - \tilde{\omega}_m$ remains large, resulting in weak field amplitudes and minimal energy localization. In contrast, high-Q resonant structures exhibit poles that lie close to the real axis. In this case, at real frequency $\omega = \mathrm{Re}[\tilde{\omega}_m]$, the term $\omega - \tilde{\omega}_m$ becomes notably small, leading to pronounced field localization and significant amplitude enhancement due to emphasis on mth resonant mode, as in below.

$$\mathbf{E}(\mathbf{r}) \approx \int \frac{\tilde{\mathbf{E}}_m(\mathbf{r}) \otimes \tilde{\mathbf{E}}_m(\mathbf{r}')}{\mathrm{Im}[\tilde{\omega}_m]} \cdot \mathbf{J}(\mathbf{r}) dv'$$

Introducing a small fictitious loss alters this behavior in a controlled manner. The additional absorption increases the imaginary part of the complex frequency $\tilde{\omega}_m$, effectively enlarging the denominator of Green's function and thereby reducing the field amplitude near resonance. Importantly, because both the frequency and modal corrections are first-order in the perturbation, such a small absorption term (on the order of 1%) primarily suppresses amplitude growth while preserving the mode shape and the real part of the resonant frequency.

From this formulation, we can further derive a heuristic guideline for selecting the material loss used in the DIRTL framework. Starting from the definition of the Q-factor associated with the QNM $\tilde{\omega}_m$, which is given by the following expression:

$$Q \equiv \mathrm{Re}[\tilde{\omega}_m]/2\mathrm{Im}[\tilde{\omega}_m]$$

under a small perturbation in permittivity $\delta\epsilon(\boldsymbol{r})$, the first-order correction to the eigenfrequency can be rewritten as

$$\delta\tilde{\omega}_m = -i \frac{\tilde{\omega}_m \int \tilde{\mathbf{E}}_m^* \cdot \delta\epsilon(\mathbf{r}) \tilde{\mathbf{E}}_m dv}{2 \int \tilde{\mathbf{E}}_m^* \cdot \epsilon(\mathbf{r}) \tilde{\mathbf{E}}_m dv} \approx -iQ\mathrm{Im}[\tilde{\omega}_m] \frac{\delta\epsilon(\mathbf{r})}{\epsilon(\mathbf{r})}$$

This leads to an increase in the imaginary part of the eigenfrequency:

$$\mathrm{Im}[\tilde{\omega}'_m] = \mathrm{Im}[\tilde{\omega}_m] \left(1 + Q \frac{\delta\epsilon(\mathbf{r})}{\epsilon(\mathbf{r})}\right)$$

Inserting this to the Green's function expansion we derived above, as field at resonance frequency scales inverse to imaginary term of quasinromal mode, by fictitious loss the field is attentuated by the factor $1 + Q\frac{\delta\epsilon(\mathbf{r})}{\epsilon(\mathbf{r})}$.

Based on this relation, the choice of fictitious loss involves a trade-off: the loss must be large enough to effectively suppress extreme resonant amplitudes, while remaining sufficiently small to preserve the modal structure under the perturbative regime. In our dataset, the maximum field amplitudes for resonant cases range approximately from 10 to 50. We therefore target an attenuation factor of $\alpha{\sim}5.$ as a practical balance between amplitude reduction and modal fidelity. This leads to

$$\frac{\delta\epsilon(\mathbf{r})}{\epsilon(\mathbf{r})} \approx \frac{\alpha - 1}{Q} \text{, or equivalently, } k \cong \frac{(\alpha - 1)n}{2Q}$$

with $\epsilon(\mathbf{r}) + i\delta\epsilon(\mathbf{r}) = (n(\mathbf{r}) + ik)^2 \approx n(\mathbf{r})^2 + i2n(\mathbf{r})k$. Using a representative maximum quality factor of $Q{\sim}400$ for our problem set, this yields $k \approx 0.01$, which is consistent with the empirically selected value used in our study. For selecting an appropriate loss factor across different problem settings, an approximate estimate of the maximum Q-factor, together with a suitably chosen attenuation factor $\alpha$, can provide a practical heuristic for determining the additional fictitious loss term $k$.

Altogether, this analysis provides a quantitative foundation for the loss-regularization strategy. It explains how the deliberate introduction of minor absorption smoothens the learning field involving high-Q resonances, effectively attenuating the high modal amplitudes without compromising the physical fidelity of the underlying modal characteristics and the position of resonant frequency. Meanwhile, it is important to note that although the derivation was done in terms of electric fields as the perturbation enters through $\delta\epsilon(\mathbf{r})$, the same attenuation mechanism also applies to the magnetic field, as the two are related through Maxwell's equation,

$$\mathbf{H} = \frac{1}{i\omega\mu_0}\nabla \times \mathbf{E}$$

## S4. Loss Tuning for Pretrain Stage

To see how much material loss is best, we demonstrated a simple sweep on different material absorption loss values. For this, we generated samples with $k = 0.01\sqrt{10}, 0.01, 0.001\sqrt{10}, 0.001$, with $n = n_{mat} + ik$, and pretrained on these to see which was best in the fine-tuned result. We pretrained and fine-tuned with respectively 1.5k structures sampled on 11 wavelength points, with pretrain stage datasets prepared on simulations with $n = n_{mat} + ik$. The f-DIRTL strategy was implemented with FNO model; 6-Layer FNO pretrained and transferred to 10-Layer fine-tuned model. The results in **Fig. S7** show that the loss we utilized; $k = 0.01$, was most effective, both in attenuating amplitudes to make a smooth learning field and preserving key mode characteristics.

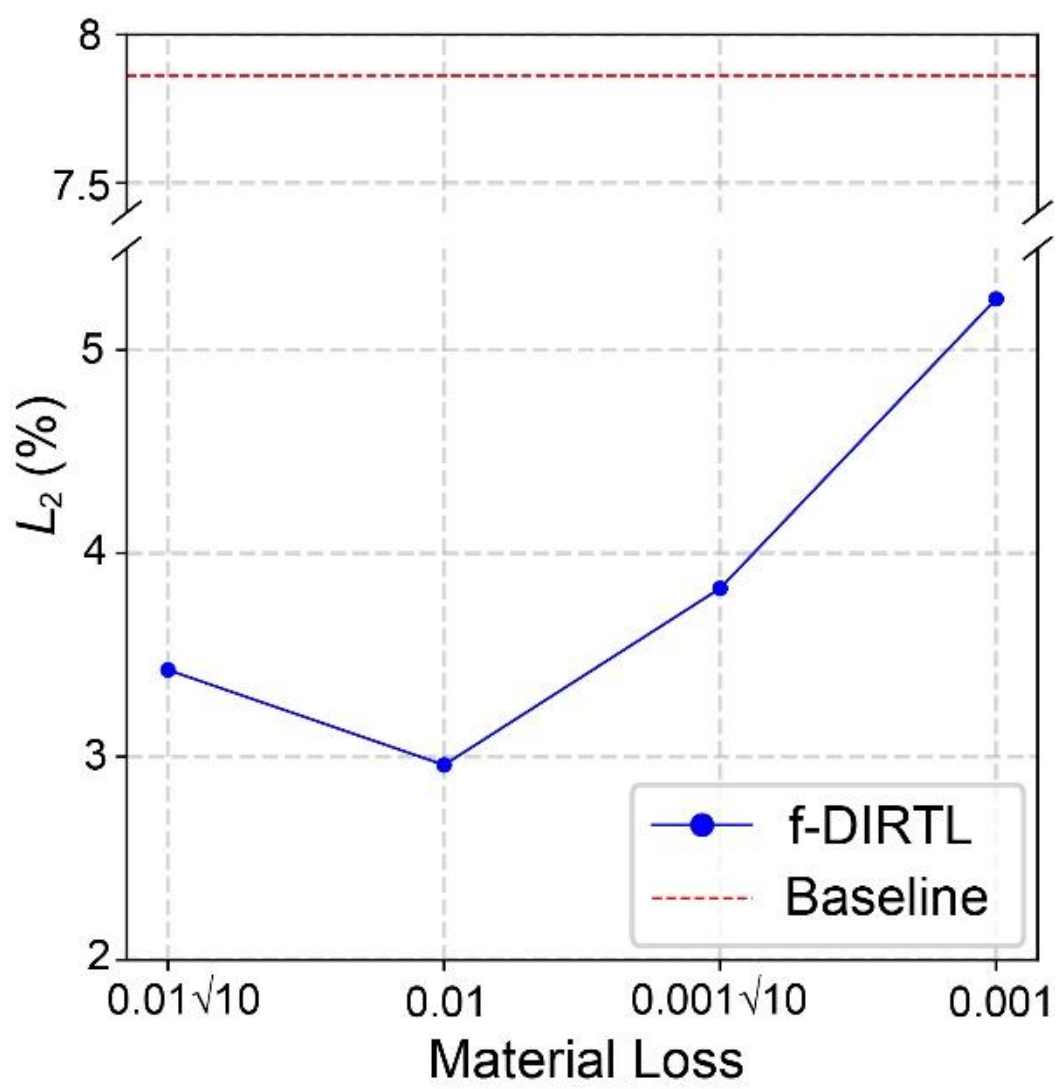

**Figure S7** Results for f-DIRTL FNO with different material loss used in the pretrain stage

## S5. Standard Curriculum and Oversampling Baseline

To further investigate the effect of data distribution on the learning of resonant EM responses, we implemented two alternative data-handling strategies: (i) dataset-duplication-based oversampling and (ii) standard curriculum learning based on resonance strength. Both approaches were constructed using the same underlying dataset and preprocessing pipeline, ensuring a fair comparison.

### Standard Curriculum Learning

To compare DIRTL to traditional curriculum learning strategies, we implemented a curriculum learning strategy where we progressively exposed the model to increasingly difficult samples based on their resonance strengths. The training dataset was partitioned into multiple stages according to predefined thresholds

- Stage 1: max $|\mathbf{H}| < 5$
- Stage 2: max $|\mathbf{H}| < 10$
- Stage 3: max $|\mathbf{H}| < 25$
- Stage 4: entire dataset

This curriculum enforces a gradual learning process, where the early stages allow the model to focus on non-resonant cases, which are easier to learn, while the later stages introduce increasingly strong resonances that the model should eventually also adapt to. The dataset distribution is shown in **Fig. S8(a)**. The training hyperparameters (FNO parameters, learning rate schedule, with step size changed to 25 to fit the curriculum learning stages) were set to be same as baseline FNO.

### Oversampling via Duplication

To address the imbalance between rare resonant samples and non-resonant cases, we implemented a strategy where we duplicate the resonant sample cases in the training dataset. For each data point, given the maximum electric field amplitude $|\mathbf{H}|$, the duplication multiplier was assigned according to logarithmic binning rule:

$$\text{multiplier} = \lceil \log_2 |\mathbf{H}| \rceil \text{ if } |\mathbf{H}| > 4, \qquad \text{else multiplier} = 1$$

This rule effectively increases the representation of high-amplitude (i.e., highly resonant) samples in the training dataset by replicating them multiple times. The resulting dataset distribution is shown in **Fig. S8(b)**. The training hyperparameters (FNO parameters, learning rate schedule) were set to be same as baseline FNO.

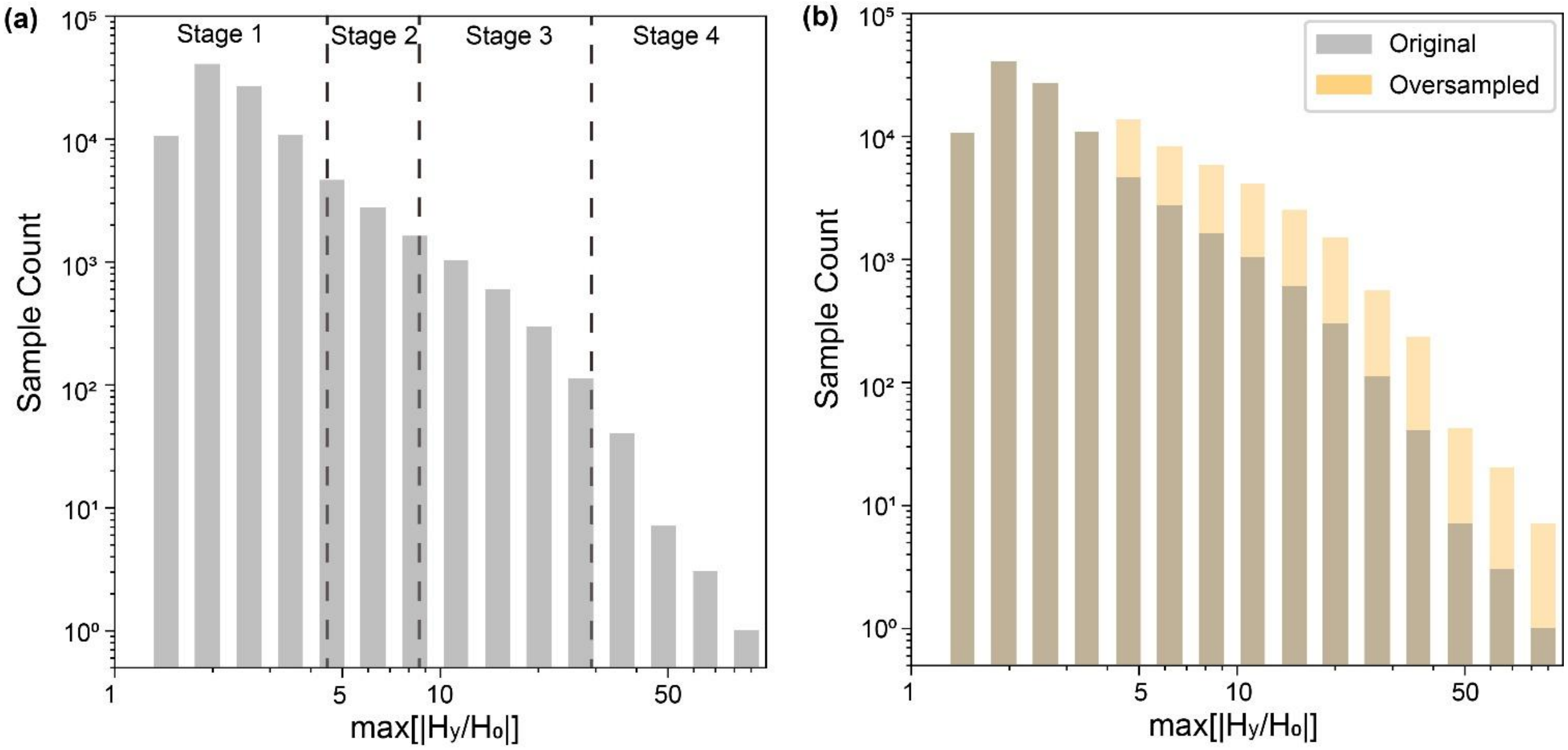

**Figure S8 Dataset distributions** for (a) standard curriculum learning implementations (b) oversampling of resonant cases.

**Table S3** Training results for various baseline trainings on FNO, UNet models.

| Model | Training Method | Normalized $L_2$(%) | Transmission Prediction Error (%) |
|---|---|---|---|
| FNO | Baseline | 5.72 | 2.49 |
| | Baseline + PINN | 5.61 | 2.56 |
| | Oversampling | 4.92 | 2.37 |
| | Standard Curriculum | 7.28 | 3.67 |
| UNet | Baseline | 9.97 | 5.40 |

| | | |
|---|---|---|
| Baseline + PINN | 10.1 | 5.40 |
| Oversampling | 10.0 | 4.65 |
| Standard Curriculum | 14.6 | 5.91 |

From **Table S3**, we can observe that a conventional curriculum learning strategy did not lead to performance improvements. This is likely because, from the model's perspective, resonant and non-resonant samples are not explicitly distinguishable from the input geometry alone, but are instead defined by their resulting field responses. Consequently, the curriculum, which was constructed based on output characteristics, does not provide a learning signal that is aligned with the input distribution, limiting its effectiveness.

Meanwhile, oversampling decreases the average test error compared to the baseline. However, this is mainly in the non-resonant region, while for the resonant cases, oversampling shows worse inference results, as we can observe in **Fig. S9(a)**. This is rather counterintuitive, as initially in our dataset design, we duplicated resonant cases so that the model would fit to that.

This behavior can be understood from two perspectives. First, resonant and non-resonant cases are not explicitly distinguishable from the input geometry alone, as the binary input appears statistically similar across both regimes. The distinction emerges only in the resulting field response. As a result, duplicating resonant cases does not introduce new informative variations in the input space, but merely repeats exposure to previously seen examples. This limitation is inevitable, as selectively generating resonant samples during data generation is not feasible in the absence of a priori knowledge of which structures will produce resonances. Meanwhile, because the model lacks an explicit input-side signal to distinguish resonant from non-resonant cases, repeated exposure does not improve its ability to identify or represent resonance-specific features [S7]. Consequently, oversampling fails to guide the model toward learning the underlying mechanisms of resonant behavior, rendering the duplication process largely redundant.

Second, the overall reduction in prediction error can be largely attributed to the increased number of effective gradient updates introduced by oversampling. Along this process, the model's inherent spectral bias becomes more pronounced: it preferentially fits smoother, low-frequency components associated with non-resonant responses, while failing to capture the sharp features of resonant fields [S8, S9]. This leads to improved accuracy in non-resonant regions but degraded performance in resonant cases, ultimately producing the opposite effect of what was intended. Also, as seen in **Fig. S9(b)**, although the transmission spectrum prediction for the oversampled case outperforms that for f-DIRTL in UNet, this is because the $L_2$ metric does not focus on the prediction of dip positions and amplitudes. The transmission dip and the overall spectrum shape is much better represented with f-DIRTL UNet.

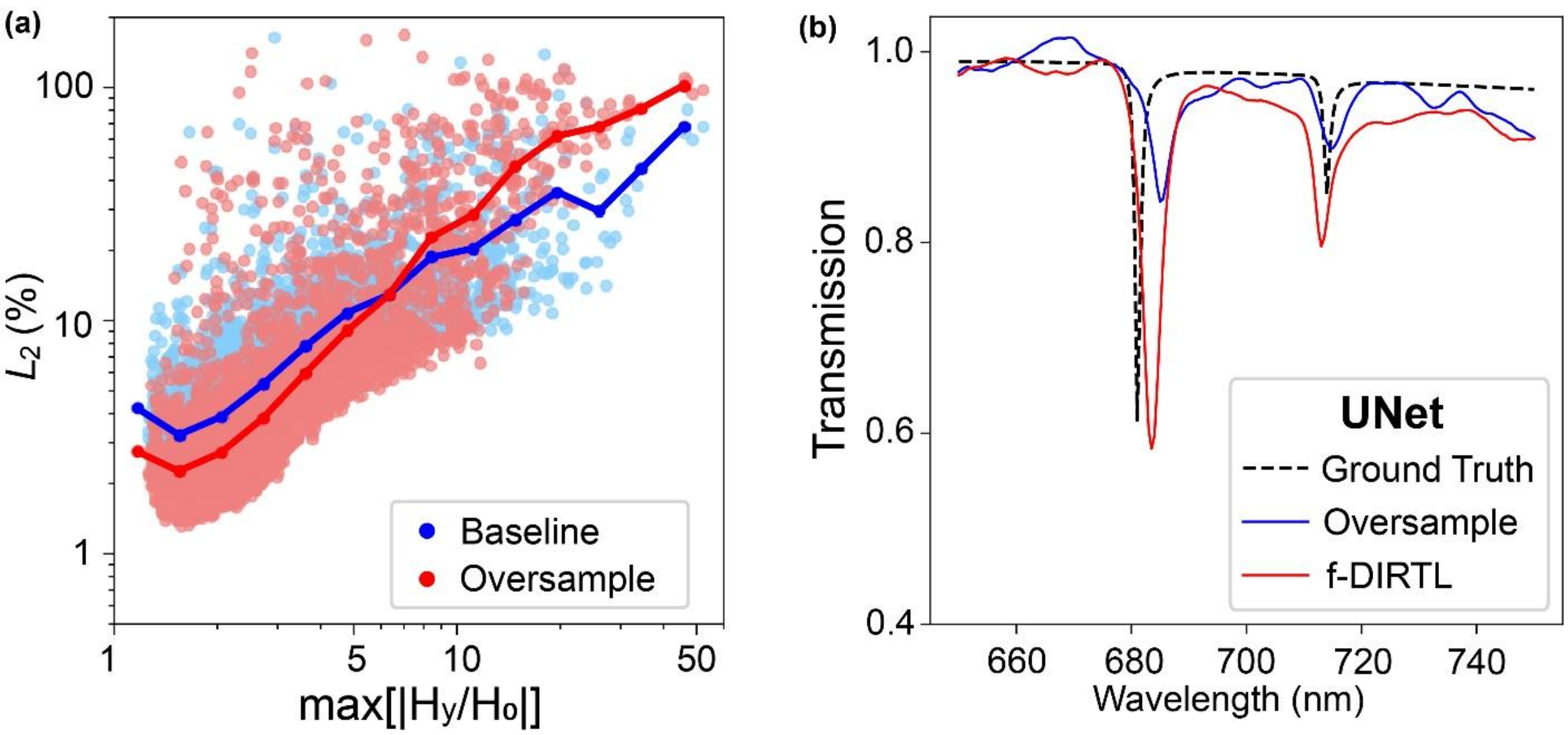

**Figure S9** (a) Error distribution plot. Comparison between baseline FNO and FNO trained on the oversampled dataset. (b) Transmission spectrum comparison between UNet trained on oversampled dataset and f-DIRTL.

## S6. Addition of Maxwell Loss

The balance between physical Maxwell loss and MSE data loss was controlled by hyperparameter α, which scales the Maxwell loss term like below.

$$L = L_{\mathrm{MSE}} + \alpha \cdot L_{\mathrm{Maxwell}}$$

$$L_{\mathrm{Maxwell}} = \frac{1}{n}\sum_{i=1}^{n} \left\| \widehat{\mathbf{H}}_y - \nabla \times \left( \frac{1}{\varepsilon(\mathbf{r})\mu_0\omega^2} \nabla \times \right) \widehat{\mathbf{H}}_y \right\|_2$$

The term α was dynamically tuned so that the ratio between $L_{\mathrm{MSE}}$ and $\alpha \cdot L_{\mathrm{Maxwell}}$ is constant. The ratio was chosen to be 0.05 for both FNO and UNet, which were all chosen via parameter sweep with learning rates identical to baseline training with only data loss [S10]. The Maxwell loss itself was implemented by discretizing the frequency-domain Maxwell equations on a Yee grid [S11].

### S7. Multi-Step DIRTL

To facilitate a smoother adaptation of the model toward learning resonant features, it is natural to adopt a multi-step learning process rather than the original two-step approach. Here, we implemented and tested multi-step DIRTL, where the material loss in the system is gradually decreased across steps. The optimizer and learning rate schedules were kept identical to those used in the two-step configuration. To ensure total data conservation, in an N-step process, the dataset for each step was set to $(\text{total number of datasets})/N$ with total 6k different structures sampled on 11 wavelength points used.

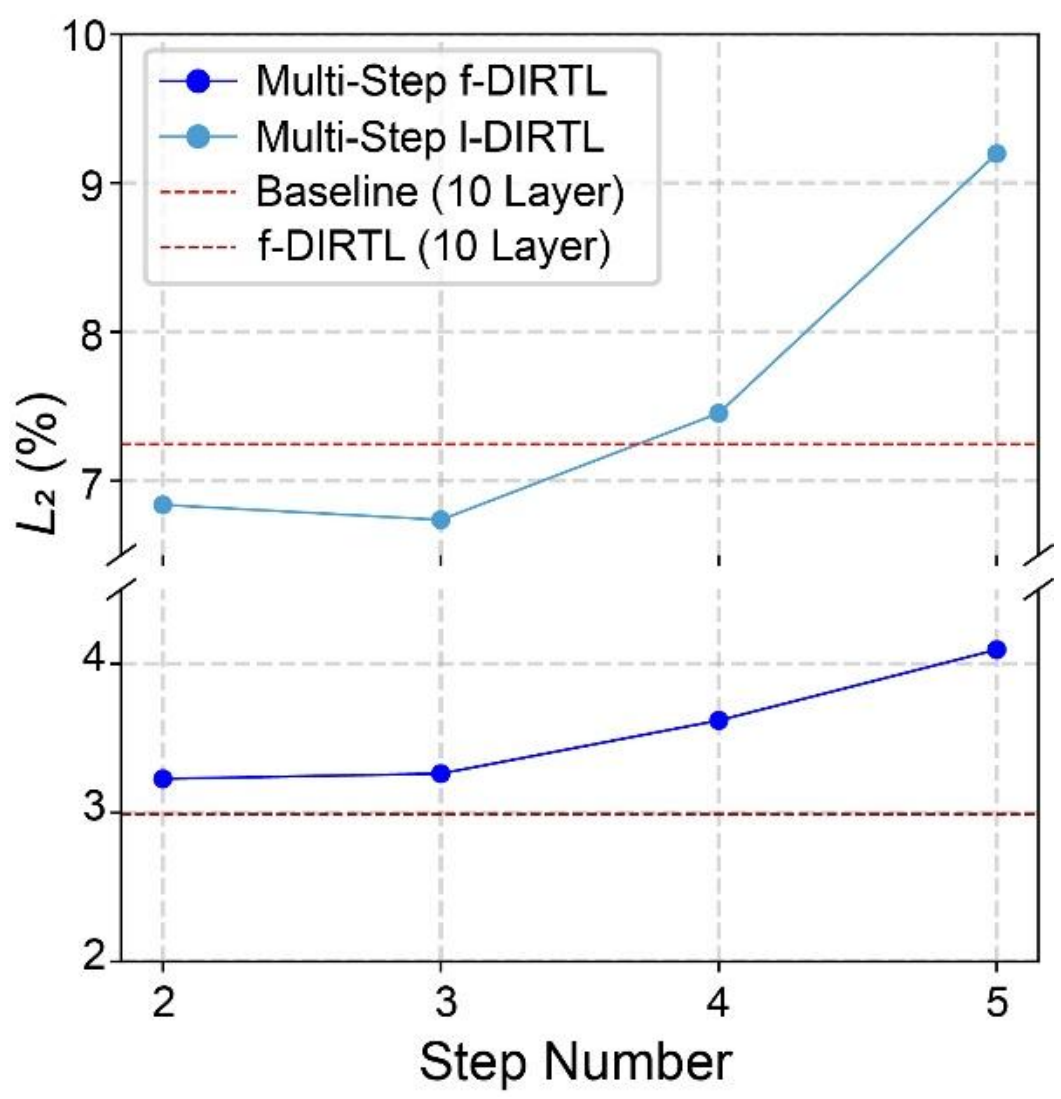

**Figure S10** Test results for multi-step DIRTL FNO, compared to baseline FNO (red dotted) and 2-step f-DIRTL FNO (6-layer pretrain, 10-layer fine-tune, dark red dotted). Dark blue plot indicates multi-step f-DIRTL, while light blue plot indicates multi-step l-DIRTL

**Table S4** Material loss schedule for each multi-step process' pretrain stage

| Step Number | Material Loss | Step Number | Material Loss |
|---|---|---|---|
| 2 Step | $[0.01]$ | 4 Step | $[0.01, 0.001\sqrt{10}, 0.001]$ |
| 3 Step | $[0.01, 0.001\sqrt{10}]$ | 5 Step | $[0.01\sqrt{10}, 0.01, 0.001\sqrt{10}, 0.001]$ |

The l-DIRTL and f-DIRTL schemes were designed as follows, with the corresponding results shown in **Fig. S10**. The $k$ values used for each n-step configuration are summarized in **Table S4**. Since $k = 0.01$ was found to be the most effective in earlier experiments, it served as the baseline for constructing the multi-step schedule. However, for the five-step process, because lower $k$ values did not produce sufficient attenuation, we began with $k = 0.01\sqrt{10}$ instead of adding smaller loss values at the end.

**l-DIRTL**

As in the two-step Loose DIRTL, the model is fine-tuned at each step using the dataset corresponding to that step, without freezing any parameters.

**f-DIRTL**

Analogous to the two-step process, after each step, additional Fourier layers are introduced and the model is fine-tuned using the dataset for that step. The training starts with four layers, and four more layers are added at each subsequent step. Consequently, an n-step process results in a model containing $4 \times N$ Fourier layers.

For l-DIRTL, we observe a modest improvement in prediction accuracy when using a three-step process, followed by a degradation in performance as additional steps are introduced. This suggests that, in principle, a multi-step relaxation schedule can provide a smoother learning trajectory for l-DIRTL. However, as the number of steps increases, the reduced number of samples allocated to each step decreases, leading to reduced prediction accuracy for each step. As result, for multi-step processes involving more than 3 steps, we see worse model capabilities. This trade-off is evident in **Fig. S10**.

For f-DIRTL, **Fig. S10** shows that accuracy consistently worsens as the number of steps increases. In this case, the loss of accuracy caused by reducing the per-step dataset outweighs any benefit from smoother progressive learning. Consequently, a two-step DIRTL procedure emerges as the optimal configuration: it offers the most substantial performance gain for f-DIRTL while maintaining accuracy comparable to the best-performing configuration for l-DIRTL.

## S8. Model Comparison for l-DIRTL

To investigate how the pretrained representations evolve during l-DIRTL training, we compare the weights of the pretrained and fine-tuned models for both FNO and UNet. For l-DIRTL FNO, both the pretrained and fine-tuned models share the same 10-layer architecture. In the case of FNO, we analyze the input and output convolution layers (lifting and projection layers), the spectral convolution within the Fourier blocks, and the linear connections inside the Fourier blocks. For UNet, we examine the convolution and pooling operations within the encoder and decoder, as well as the input and output convolution layers. The similarity between pretrained and fine-tuned weights is quantified using cosine similarity, which is defined as follows [S3].

$$\text{cosSim}(w_1, w_2) = \frac{w_1 \cdot w_2}{\|w_1\|_2 \|w_2\|_2}$$

**Figure S11(a–c)** shows that, for the FNO model, the cosine similarity between the pretrained and l-DIRTL fine-tuned weights lies in the range of 0.9 to 1.0 across layers. This indicates that a significant portion of the pretrained information is retained during fine-tuning. However, comparison with the UNet results in **Fig. S11(d–f)** reveals a notable difference: UNet exhibits cosine similarity values extremely close to 1.0 across almost all layers, suggesting minimal deviation from the pretrained parameters.

This contrast is speculated to imply that the parameter updates in FNO are not merely negligible perturbations, but rather reflect a more global adaptation of the model to the new target distribution [S12]. In contrast, for the case of UNet, the encoder–decoder architecture provides a natural structural separation of roles: the encoder largely preserves pretrained representations, while the decoder adapts to the new task, as shown in how the parameters are deviated from the pretrained model [S13]. This division is predicted to enable stable retention of learned features while allowing flexible adaptation during fine-tuning.

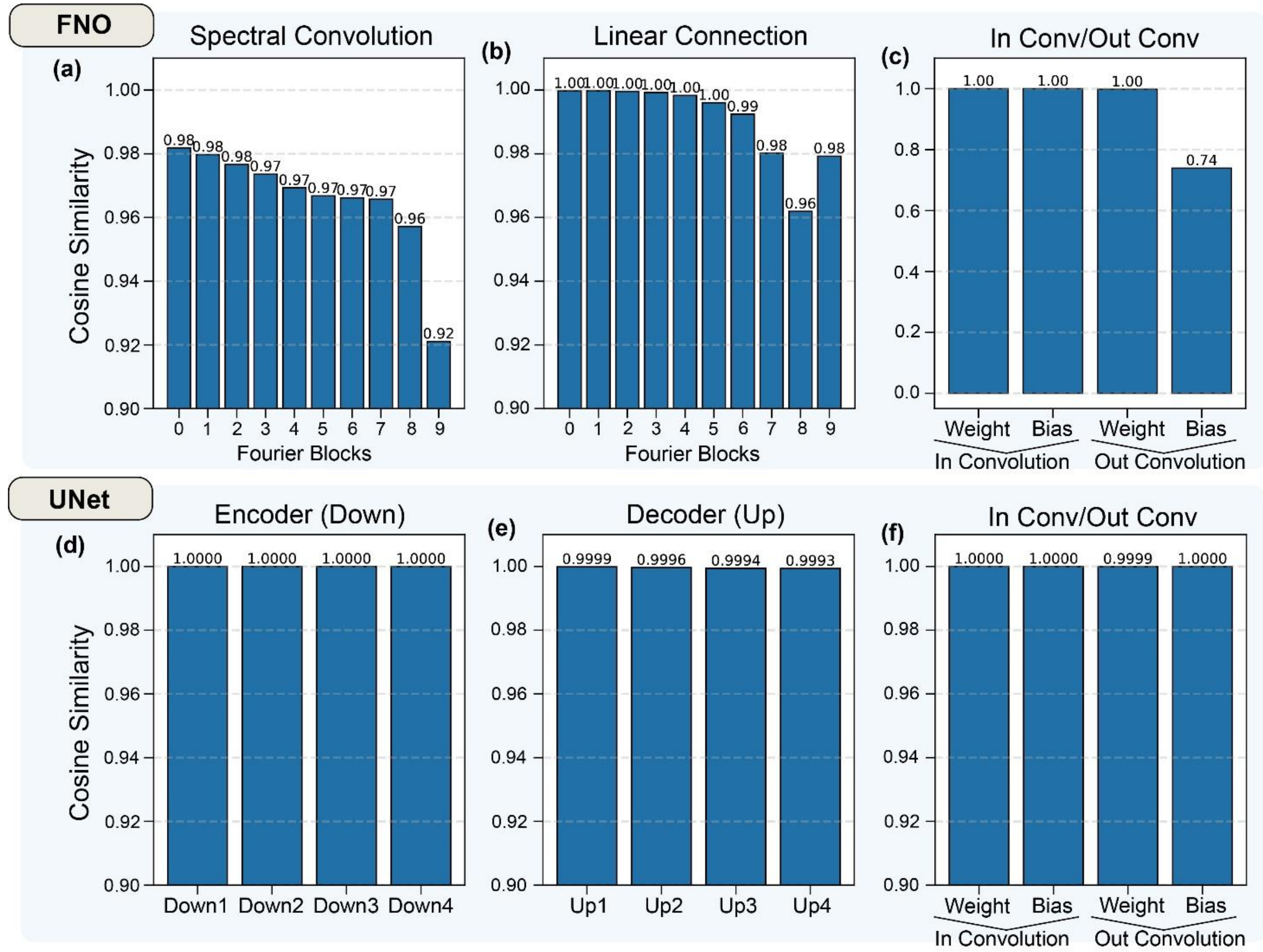

**Figure S11 Cosine similarity between pretrained and fine-tuned model for l-DIRTL FNO and UNet.** Parameter similarities for FNO. Similarities for (a) spectral convolution parameters (b) linear connections inside Fourier blocks. (c) Similarities for lift and projection convolutional layers. Parameter similarities for UNet. (a) Downscaling encoder layers. (b) Upscaling decoder layers. (c) Lift and projection convolutional layers.

In contrast, FNO lacks such structural separation. The learned representations are distributed across the Fourier layers, which collectively define the operator. As a result, fine-tuning introduces small but globally distributed parameter updates throughout the network. Although each update is minor, its cumulative effect alters how the pretrained information is utilized, effectively reshaping the learned operator. Consequently, it is expected that the pretrained knowledge in FNO becomes more volatile during the l-DIRTL process, as it is not structurally preserved within specific subnetworks. This lack of structural preservation likely contributes to the reduced data efficiency observed for l-DIRTL in FNO compared to UNet. This logic also naturally motivates the adoption of parameter-freezing strategies, which help preserve pretrained information during fine-tuning and stabilize the transfer process.

## S9. Multi-Task Performance of f-DIRTL FNO

To evaluate the generalizability of the proposed training strategy, we compare the predictive performance of the baseline FNO and f-DIRTL FNO on wavelengths during training. We generated 1,000 test samples uniformly sampled at 0.5 nm intervals. As shown in Supporting Fig. 6, f-DIRTL FNO consistently achieves higher accuracy across the entire wavelength domain. Furthermore, baseline FNO exhibits a pronounced increase in error near ~700 nm, where most high-resonant cases occur. Meanwhile, f-DIRTL FNO suppresses this error curvature, demonstrating reliable performance even in resonance-dominated regimes where prediction is typically difficult.

The pretrained core also shows strong multi-tasking capabilities in both wavelength interpolation and extrapolation. In the interpolation setting, the model is first trained on a coarsely sampled wavelength grid (20 nm interval within 650–750 nm) and later fine-tuned on a denser grid (10 nm interval), using 4.5k structures for each stage while following f-DIRTL FNO procedure. This entire process uses fewer training samples than the baseline model, which is trained directly on the dense wavelength grid with 9k structures. As shown in **Fig. S12**, the interpolated f-DIRTL FNO achieves substantially lower prediction error across both trained and untrained wavelengths, demonstrating that the pretrained core encodes functionally meaningful wavelength-response relationships.

More importantly, the pretrained core enables effective extrapolation, a significantly more challenging task for NNs [S14, S15] The compact model is first pretrained over the limited 680–720 nm region and then fine-tuned over the full 650–750 nm band, using 4.5k structures for each stage while following f-DIRTL FNO procedure. Even with fewer training samples than the baseline, the resulting model achieves higher accuracy in the previously unseen 650–680 nm region and comparable or improved accuracy in the 720–750 nm region, as shown in **Fig. S12**. This demonstrates that the pretrained core forms wavelength representations that remain stable under spectral extension, enabling robust extrapolation.

Overall, these results indicate that the pretraining phase in f-DIRTL establishes a physically informed core representation that enables robust generalization across multiple wavelength inputs. This core representation can be directly extended to related tasks, including wavelength interpolation and extrapolation, while preserving the data-efficiency of the approach. Although our demonstration focuses on multi-wavelength prediction, the same pretrained core has the potential to support broader multi-task learning scenarios—such as variations in incident angle, device geometry, or material refractive index—while still maintaining high training efficiency.

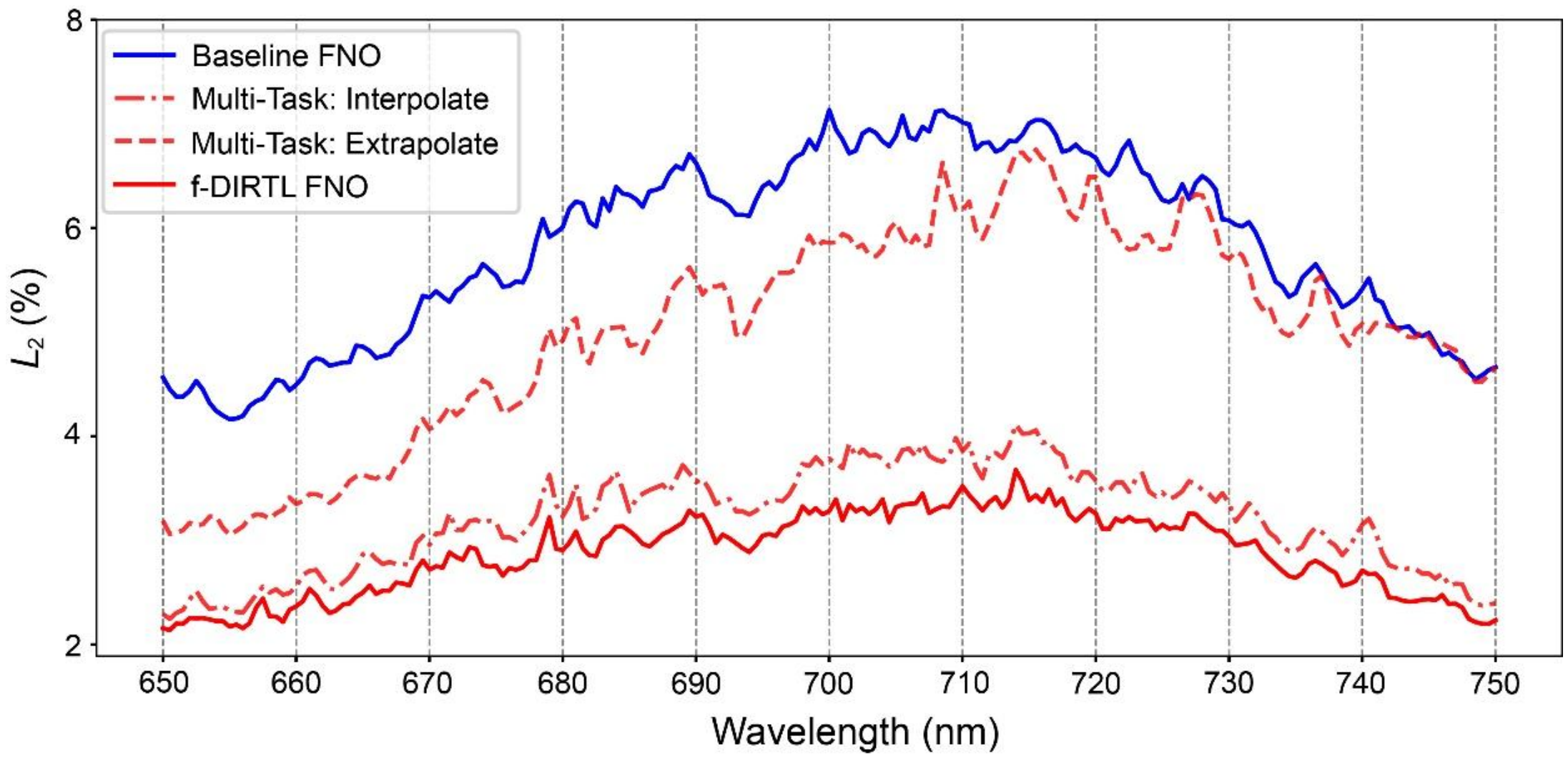

**Figure S12 Multi-tasking evaluation of the f-DIRTL FNO.** This demonstrates comparison between the baseline FNO and f-DIRTL FNO on wavelength points not included in training. Gray dotted lines denote the 10nm training wavelengths; testing is performed at 0.5nm resolution across 1,000 structures. For interpolation multi-tasking case, the FNO model is pretrained 20nm interval wavelengths and fine-tuned on all 10nm training points. For extrapolation multi-tasking case, the f-DIRTL model is pretrained on the wavelengths 680nm ~ 720nm with 10nm intervals and fine-tuned on all 10nm training points.

## S10. Visualization of Learning Dynamics of f-DIRTL FNO

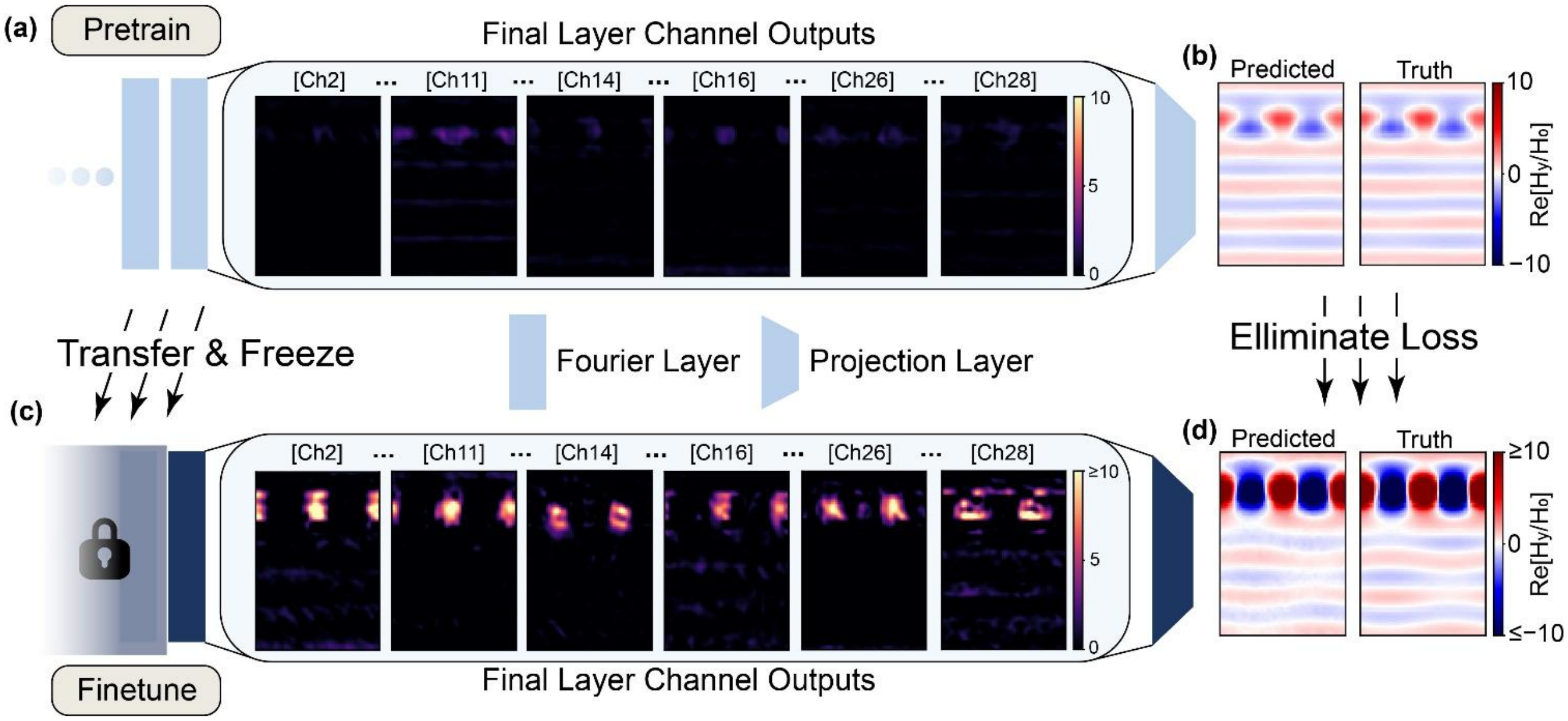

**Figure S13 Learning dynamics of f-DIRTL FNO** (a) Visualization of the final Fourier layer channel outputs during the pretraining stage using lossy data. The feature maps exhibit attenuated field patterns, reflecting the smoothed resonant responses learned under dissipative conditions. (b) Corresponding field prediction from the pretrained model, showing strong agreement with the ground truth for attenuated resonant fields. (c) Final-layer channel of the fine-tuned model on the lossless dataset. The feature maps evolve to exhibit sharper and higher-amplitude resonant characteristics, indicating adaptation to the true high-Q response. (d) Final prediction after fine-tuning, demonstrating accurate reconstruction of high-amplitude resonant fields once the fictitious loss is removed.

**Figure S13** illustrates the overall learning dynamics of the f-DIRTL FNO. As shown in **Fig. S13(b)**, the pretrained model accurately predicts the attenuated resonant field distributions, indicating that the network has successfully captured the underlying global modal structure under the smoothed (lossy) training regime. This well-initialized representation subsequently enables effective inference on the target lossless dataset, as demonstrated in **Fig. S13(d)**, where high-amplitude resonant features are accurately reconstructed after fine-tuning.

The internal learning dynamics are further revealed in **Fig. S13(a, c)**, which visualize the outputs of the final Fourier layers during the pretraining and fine-tuning stages, respectively. In the pretrained model, the final-layer channel outputs already resemble the attenuated field distributions, suggesting that the network has formed a physically meaningful representation of the global field response before the projection layer. After fine-tuning, the corresponding feature maps evolve to exhibit high-amplitude resonant patterns, indicating that the deeper layers adapt to recover sharp resonant characteristics based on the information from pretrained layers while preserving the global structure.

## S11. DIRTL on Lossy Target Data

On the topic of real-world applicability, it is important to consider nanophotonic devices composed of dispersive media, where the constituent materials exhibit nonzero loss. In such systems, the target field data already show partially attenuated resonant responses because material absorption suppresses the resonance amplitude. This naturally raises the question of whether DIRTL remains beneficial in these cases, since the contrast between resonant and non-resonant samples is less extreme than in the ideal lossless setting. To address this point, we evaluated DIRTL on datasets in which the target data themselves included material loss.

The dataset configuration was kept the same as in the previous analysis, namely a 1D binary grating with refractive index $n = 2$, while only the material loss was varied. Specifically, we tested three target-loss conditions, $k = 0.01, 0.001\sqrt{10}(\approx 0.0032), 0.001$, using the same model architecture and hyperparameters as those employed for the lossless target-data case. For the baseline FNO model consisting of 10 Fourier layers, training was conducted using 3,000 randomly generated structures. For the DIRTL comparison, a two-stage training procedure was adopted. First, a 6-layer FNO model was pretrained on 1,500 randomly generated structures with material loss (lossy dataset). The pretrained weights were then transferred to a 10-layer FNO model, where the transferred layers were frozen. The model was subsequently fine-tuned on the target dataset of 1,500 randomly generated structures, enabling a direct comparison with the baseline under an equivalent total data budget.

For the pretraining stage, the material loss was chosen to be ten times larger than that of the corresponding target dataset. Specifically, for target losses of $k = 0.01, 0.001\sqrt{10}(\approx 0.0032), 0.001$, the pretraining datasets were generated with losses of $k = 0.1, 0.1\sqrt{10}(\approx 0.0316), 0.01$. This choice was made to ensure sufficient additional attenuation even for already lossy target cases, while maintaining a consistent scaling relationship across different target-loss conditions. Using the attenuation relation we derived above

$$\frac{\delta\epsilon(\mathbf{r})}{\epsilon(\mathbf{r})} \approx \frac{\alpha - 1}{Q}$$

this should lead to about 10 times attenuation. However, as perturbation theory breaks down for larger losses, we expect less amplitude reduction effects.

Results presented in **Table S5** and **Fig. S14** clearly demonstrate that DIRTL remains effective for cases with small to moderate material loss ($k = 0.001, 0.0032$), while its performance degrades for highly lossy target datasets ($k = 0.01$). This indicates that DIRTL retains its advantage in the presence of minuscule loss, provided that the system still supports sufficiently high-Q resonances. In such regimes, the introduction of a small fictitious loss can broaden the resonance linewidth without significantly shifting the resonant frequency, thereby preserving the underlying modal characteristics. This enables the pretraining stage to effectively capture the global modal features, leading to improved performance during fine-tuning.

Meanwhile, for systems without such high-Q resonances, introducing a small amount of fictitious loss does not significantly attenuate the field amplitudes, rendering the pretraining stage less effective and potentially diminishing the data-efficiency advantage of DIRTL compared to the baseline. In such low-Q regimes, achieving meaningful attenuation would require a relatively large additional loss. However, under these conditions, the perturbative assumption underlying the DIRTL framework breaks down, and the resonance frequencies may shift appreciably. This can distort the underlying physical response, thereby reducing the effectiveness of the pretrained representations and limiting the benefits of the DIRTL strategy.

**Table S5** Training results for baseline FNO and f-DIRTL FNO for different target datasets

| Target Dataset | Normalized $L_1$(%) | | Normalized $L_2$(%) | |
|---|---|---|---|---|
| | Baseline | f-DIRTL | Baseline | f-DIRTL |
| $k = 0.01$ | 2.20 | 2.30 | 2.73 | 3.14 |
| $k = 0.001\sqrt{10}(\approx 0.0032)$ | 3.31 | 2.34 | 4.06 | 3.18 |
| $k = 0.001$ | 4.27 | 2.50 | 5.27 | 3.35 |

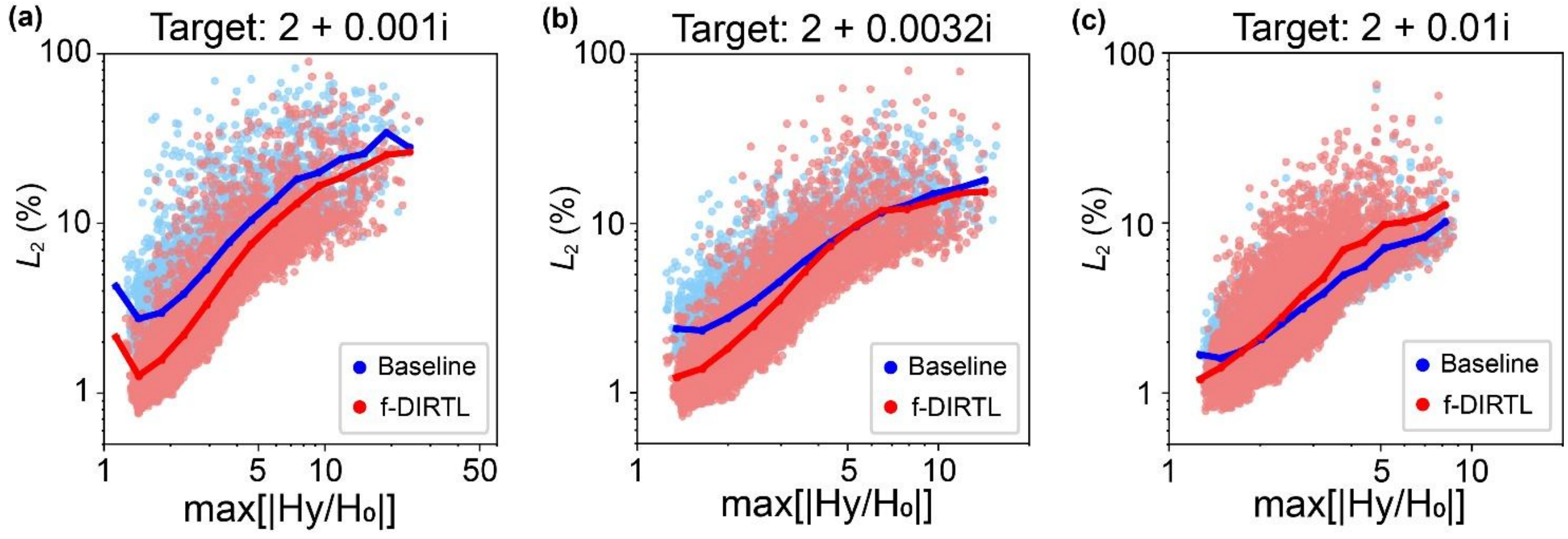

**Figure S14 Error distribution comparison plot between baseline FNO and f-DIRTL FNO for different target datasets**. Each light-blue (baseline FNO) and light-coral (f-DIRTL FNO) point represents the prediction error of an individual test sample plotted against its maximum field amplitude (a) Results for target dataset generated with material loss $k = 0.001$. (b) Results for target dataset generated with material loss $k = 0.0032$. (c) Results for target dataset generated with material loss $k = 0.01$.

## S12. Generalization to General Device Structures

While the preceding sections focus on the 1D binary grating as a controlled benchmark, a critical next step is to examine how the DIRTL framework extends to more realistic EM problems. In practical nanophotonic applications, the governing physics becomes significantly more complex. These higher-dimensional field distributions, multiple interacting resonances, and sensitivity to fabrication variations. These characteristics increase the difficulty of accurate field prediction not only for resonant field cases but also in the global picture, as predicting the overall field scheme itself becomes challenging. In this context, it becomes essential to address the real-world applicability of DIRTL beyond simplified benchmark systems. The key question is not merely whether the method performs well under idealized conditions, but whether its underlying training strategy can provide consistent benefits when applied to practical device geometries and broader parameter spaces.

### 2D Multilayer Device

To further evaluate the generality of DIRTL, we apply the framework to a different two-dimensional layered structure. In this case, the device consists of a binary random pattern in both the x and y directions. Specifically, the structure is composed of 8 cells along the x-axis and 4 cells along the y-axis, where each cell is randomly assigned either $SiO_2$ ($n = 1.46$) or $Si_3N_4$ ($n = 2.02$), with operation wavelengths centered around 500 nm. The periodicity is fixed at 1μm, and each layer has a thickness of 1μm, resulting in a total device thickness of 4μm. The transmission medium is set to $SiO_2$. Training data were generated under TM-polarized plane-wave incidence at 3 wavelength points (500, 550, and 600 nm). The overall simulation setup is illustrated in **Fig. S15(a)**.

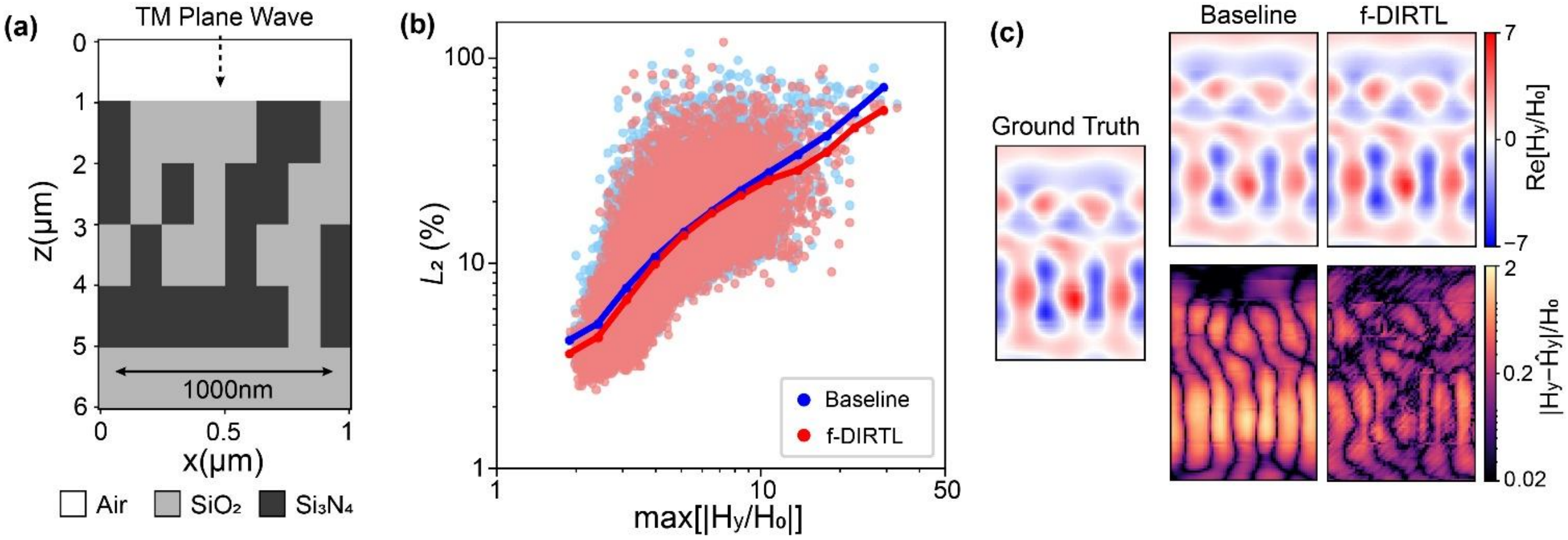

**Figure S15 Problem schematic and results of 2D multilayer device problem.** (a) Schematic of the 2D multilayer problem setup. The design space consists of random binary patterns of SiO2 and Si3N4 on a SiO2 substrate. The structure comprises four stacked layers along the z-direction, each containing 8 cells along the x-direction. (b) Error distribution comparison between the baseline FNO and f-DIRTL FNO on test samples. Both models were trained on a dataset of 320k samples and evaluated on a test set of 10k randomly generated structures. (c) Comparison of predicted field distributions for the baseline FNO and f-DIRTL FNO (top), along with the corresponding error maps (bottom).

**Table S6** Training results of 2D multilayer device. Comparison of baseline FNO and f-DIRTL FNO trained on each dataset size. The dataset size indicates how many random structures were used, while each structure was sampled on 3 wavelength points. Comparison is done via normalized $L_2(\%)$ error.

| Training Method | Error for Each Training Dataset Size (%) | | | | | |
|---|---|---|---|---|---|---|
| | 10k | 20k | 40k | 80k | 160k | 320k |
| Baseline FNO | 28.8 | 23.6 | 18.4 | 14.8 | 12.7 | 10.8 |
| f-DIRTL FNO | 31.7 | 23.7 | 18.5 | 14.5 | 12.1 | 10.0 |

For comparison between the baseline and f-DIRTL approaches, we employed FNO model. The pretrained model consists of 8 layers, while the final fine-tuned model uses 10 layers. To ensure sufficient representational capacity during pretraining, a relatively larger model was used for pretrain, and the first 5 layers were subsequently transferred to the fine-tuning stage, with the remaining layers left trainable to enable effective adaptation to the target task. To ensure a fair comparison, hyperparameters, including learning rates and model architectures, were optimized for both baseline and f-DIRTL. For the DIRTL framework, the total dataset was evenly divided between the pretraining and fine-tuning stages. The pretraining dataset was generated with artificial material loss, while the fine-tuning dataset was lossless. Specifically, for the lossy dataset, a material loss of $k = 0.01$ was introduced for both material indices, consistent with the 1D binary grating problem.

As shown in **Fig. S15(b,c)** and **Table S6**, the f-DIRTL model outperforms the baseline under sufficient training data. Notably, at smaller dataset sizes, the baseline exhibits better inference performance than f-DIRTL. However, as the amount of training data increases, f-DIRTL surpasses the baseline, demonstrating superior scalability with respect to data. This behavior can be attributed to the increasing accuracy of the pretrained model as the dataset grows, which directly enhances the quality of the fine-tuned model. These results highlight the importance of a well-trained pretrained core in the DIRTL framework. Importantly, this requirement does not limit the practicality of the method. In typical surrogate-solver applications within downstream optimization workflows, models are already trained on large datasets to achieve high accuracy. Incorporating DIRTL into such pipelines can further improve data efficiency while enhancing predictive performance.

From a theoretical perspective, the underlying mechanism of DIRTL is not restricted to 1D systems. The framework is motivated by the relaxation of high-Q resonances through fictitious loss, grounded in QNM physics. This mechanism is general and can naturally extend to higher-dimensional systems, where resonant behavior likewise originates from QNM dynamics, albeit with more complex modal coupling, polarization mixing, and interference effects. Consistent with this expectation, DIRTL also demonstrated performance improvements for the considered 2D multilayer problem, suggesting that the framework has the potential to generalize beyond the simple 1D binary grating dataset explored in this work.

Meanwhile, we observe a degradation in the performance of DIRTL compared to the 1D binary problem presented in the main text. This can be attributed to the increased complexity of the global field dynamics in the present problem, where the model must capture not only resonant behavior but also the overall field distribution. As a result, the model struggles to accurately represent the full physical response. This challenge is particularly pronounced in low-data regimes, where inaccuracies in the pretraining stage lead to significant misrepresentation that propagates into the fine-tuning phase, as discussed previously.

These results suggest that, for more complex and higher-dimensional problems, addressing resonance alone is insufficient for DIRTL to succeed. Accurately capturing the global field response remains a critical challenge. Therefore, future progress will require not only improvements in handling resonant features but also advances in dataset construction, model architecture design, physics-informed learning approaches, and transfer learning strategies.